\documentclass{aa}  

\usepackage{graphicx}
\usepackage{txfonts}
\usepackage{graphicx}
\usepackage{mathrsfs}
\usepackage{url}
\begin{document}

\title{Chains of dense cores in the Taurus L1495/B213 complex
\thanks{Based on observations carried out with the IRAM 30m Telescope. 
IRAM is supported by INSU/CNRS (France), MPG (Germany), and IGN (Spain).}
\thanks{{\it Herschel} is an ESA space observatory with science instruments provided by European-led Principal Investigator consortia and with important participation from NASA.}}

\author{M. Tafalla \inst{1}
\and
A. Hacar \inst{2}
}

\institute{Observatorio Astron\'omico Nacional (IGN), Alfonso XII 3,
E-28014 Madrid, Spain
\email{m.tafalla@oan.es}
\and
Institute for Astrophysics, University of Vienna,
T\"urkenschanzstrasse 17, A-1180 Vienna, Austria\\
\email{alvaro.hacar@univie.ac.at}
}

\date{}

 \abstract
   {Cloud fragmentation into dense cores 
   is a critical step in the process of star formation. 
   A number of recent observations show that it is 
   connected to the filamentary structure of the gas, but the
   processes responsible for core formation remain mysterious.}
   {We studied the kinematics and spatial distribution of the dense gas
   in the L1495/B213 filamentary region of the Taurus molecular cloud
   with the goal of understanding the mechanism of core formation. }
  {We mapped the densest regions of L1495/B213 in
   N$_2$H$^+$(1--0) and C$^{18}$O(2--1) with the IRAM 30m telescope, 
   and complemented
   these data with archival dust-continuum observations from the Herschel Space 
   Observatory.}
   {The dense cores in  L1495/B213 are significantly clustered in 
   linear chain-like groups about 0.5~pc long.
   The internal motions in these chains
   are mostly subsonic and the velocity is continuous, 
   indicating that turbulence dissipation
   in the cloud has occurred at the
   scale of the chains and not at the smaller scale of the individual cores.
   The chains also present an approximately constant abundance of 
   N$_2$H$^+$ and radial intensity profiles that can be modeled
   with a density law that follows a softened power law.
   A simple analysis of the spacing between the cores 
   using an isothermal cylinder model
   indicates that the cores have likely formed by gravitational
   fragmentation of velocity-coherent filaments.
   }
   {Combining our analysis of the cores with our previous study of
   the large-scale C$^{18}$O emission from the cloud, 
   we propose a two-step scenario of core formation in L1495/B213. 
   In this scenario, named {\em ``fray and fragment''},
   L1495/B213 originated from the supersonic collision of two flows.
   The collision produced a
   network of intertwined subsonic filaments or fibers ({\em fray step}).
   Some of these fibers  
   accumulated enough
   mass to become gravitationally unstable and 
   {\em fragment} into chains of closely-spaced cores.}


   \maketitle
%

\section{Introduction}

Star formation requires a high degree of cloud fragmentation.
A typical dark cloud is tens of parsecs in size,
but the cores
that undergo gravitational collapse and form stars are 
less than 0.1~pc in diameter. Understanding how 
a large-scale cloud of gas fragments into a small number
of dense cores remains a critical challenge
in the field of star formation 
\citep{dif07,war07,ber07}.

A clue to understanding fragmentation comes from cloud morphology.
Molecular clouds are known to present complex
filamentary distributions
over multiple size scales, and a connection between
this filamentary structure and the process of cloud fragmentation 
has long been proposed \citep{sch79,lar85,har02,mye09}.
Interest on this connection has been boosted by 
the large-scale cloud images from the Herschel Space 
Observatory, which display 
a striking prevalence of filamentary structures
in the distribution of cloud material \citep{and10,mol10}.
These new Herschel images show that dense cores often lie along
large-scale filaments like beads in a string, and 
leave little doubt that some type of filamentary fragmentation 
must be responsible for their condensation (see
\citealt{and13} for a recent review).

While filamentary fragmentation appears to produce cores,
the exact manner in which this process operates is far from clear.
Filamentary structures are often as large as the clouds themselves
and involve most of the cloud mass, but core and star formation
have an efficiency of only a few percent \citep{eva09}.
Filaments therefore cannot completely fragment into cores,
and some process must prevent most mass in a filament 
to end up forming cores and stars. What limits fragmentation is still a
mystery, especially considering that many of the observed
filaments 
have estimated mass-per-unit-lengths that greatly exceed
the limit of gravitational stability \citep{arz11,hen12,pal13}

In a previous study of the filamentary region L1495/B213 in
Taurus \citep{hac13}, we used velocity information 
derived from the C$^{18}$O emission to decompose
what looks like a single filament in
optical and continuum images
into a complex network of 35 smaller filamentary structures.
These structures, referred to as ``fibers'' to
distinguish them from the large-scale filament,
present properties that differ significantly from those 
of the 10~pc-long L1495/B213 region. The fibers, for example,
have typical sizes around 0.5~pc, coherent velocity fields,
and mass-per-unit-lengths that
lie within uncertainties in the expected range
of gravitational equilibrium values. 
These fibers seem to represent a size scale intermediate between the
large filamentary cloud and the smaller dense cores, and
have likely formed by some type of fragmentation process
associated with the dissipation of
turbulence  \citep{hac13}.

When the C$^{18}$O data of L1495/B213 were complemented with N$_2$H$^+$
observations, which highlight the dense cores, the fibers were
found to divide into two groups. Most fibers did not contain
embedded cores, and were referred to as ``sterile,'' but a
small group of fibers contained the totality of the cores 
and were classified as ``fertile.'' 
This difference between sterile-fertile fibers was significant.
Sterile fibers did not contain cores, but fertile
fibers contained around three cores on average.
As a result, most cores in the L1495/B213 were found
to be located
in characteristic closely-packed linear groups.

The low angular resolution observations of \citet{hac13},
made with the 14m FCRAO telescope, limited the study 
of the closely-packed cores to only the most basic global properties.
To remedy this,  we carried out higher-resolution observations 
using the IRAM 30m telescope.
These new observations allow us 
to resolve the internal structure of the core linear groups and to study the
connection between the different cores formed from a single fiber.
In the following sections we present the analysis of the core
emission with emphasis on the kinematics of the gas. In the
last section, we present a simple scenario of core formation 
that combines the large-scale
analysis of \citet{hac13} with the results from the new 
IRAM 30m data.

\section{Observations}
We observed selected regions of the L1495/B213 cloud 
with the IRAM 30m telescope 
during one session in February-March 2013 and 
another one the following June. In both sessions we 
used the EMIR heterodyne receiver \citep{car12}
in frequency-switching mode 
together with the VESPA autocorrelator.

The observations consisted of simultaneous
on-the-fly maps 
in the lines of N$_2$H$^+$(1--0) (93.17~GHz)
and C$^{18}$O(2--1) (219.56 GHz)
in dual polarization mode.
The maps covered the regions
identified by \citet{hac13} as
bright in N$_2$H$^+$(1--0) and therefore
indicative of dense core formation.
To resolve the lines in velocity, 
the VESPA autocorrelator was set
to a frequency resolution of 
20~kHz, which corresponds to 0.063~km~s$^{-1}$
at the frequency of N$_2$H$^+$(1--0)
and to 0.027~km~s$^{-1}$ at the frequency of C$^{18}$O(2--1)

The data were calibrated by observing
a combination of ambient and cold loads plus
the blank sky
every 10 minutes approximately.
The resulting $T_\mathrm{A}^*$ scale 
was converted into main beam
brightness temperature $T_\mathrm{mb}$ using the facility-recommended
main beam efficiencies of 0.8 and
0.6 for N$_2$H$^+$ and C$^{18}$O, respectively.
All intensities in this paper
are reported in $T_\mathrm{mb}$ scale and
have an estimated uncertainty of 10-15\%.

Additional off-line data processing  was carried out using the
GILDAS program CLASS\footnote{\url{http://www.iram.fr/IRAMFR/GILDAS}},
and included convolving the data with
a Gaussian kernel to resample the observations into a regular spatial 
grid \citep{man07}, folding the spectra
to correct for frequency switching, and subtracting 
a polynomial to flatten the baseline.
In some steps of the analysis, the data, which have an
original angular resolution of $26''$ and $12''$ for N$_2$H$^+$ and C$^{18}$O,
were further convolved with a Gaussian of full with half maximum (FWHM)
of $20''$ to enhance the sensitivity. 
These convolved data have an angular resolution of $33''$ (N$_2$H$^+$) and 
$23''$ (C$^{18}$O).

To complement the IRAM 30m observations, we used archival data from the 
Herschel Space Observatory \citep{pil10}. These data
consisted of dust continuum maps of the L1495/B213 region 
observed with the SPIRE instrument 
at 250, 350, and 500 $\mu$m \citep{grif10}. They were 
obtained as 
part of the Herschel Gould Belt Survey (HGBS, \citealt{and10}),
and have been previously presented
by \citet{pal13} and \citet{kir13}. 
The data used here (OBSID 1342202254) were obtained
through the Herschel Science
Archive and correspond to level 2.5 as reduced with 
version 9.1.0 of the Standard Product Generation (SPG) software.

\section{Results}

\subsection{Large-scale distribution of the gas in L1495/B213}

\begin{figure}
\resizebox{\hsize}{!}{\includegraphics{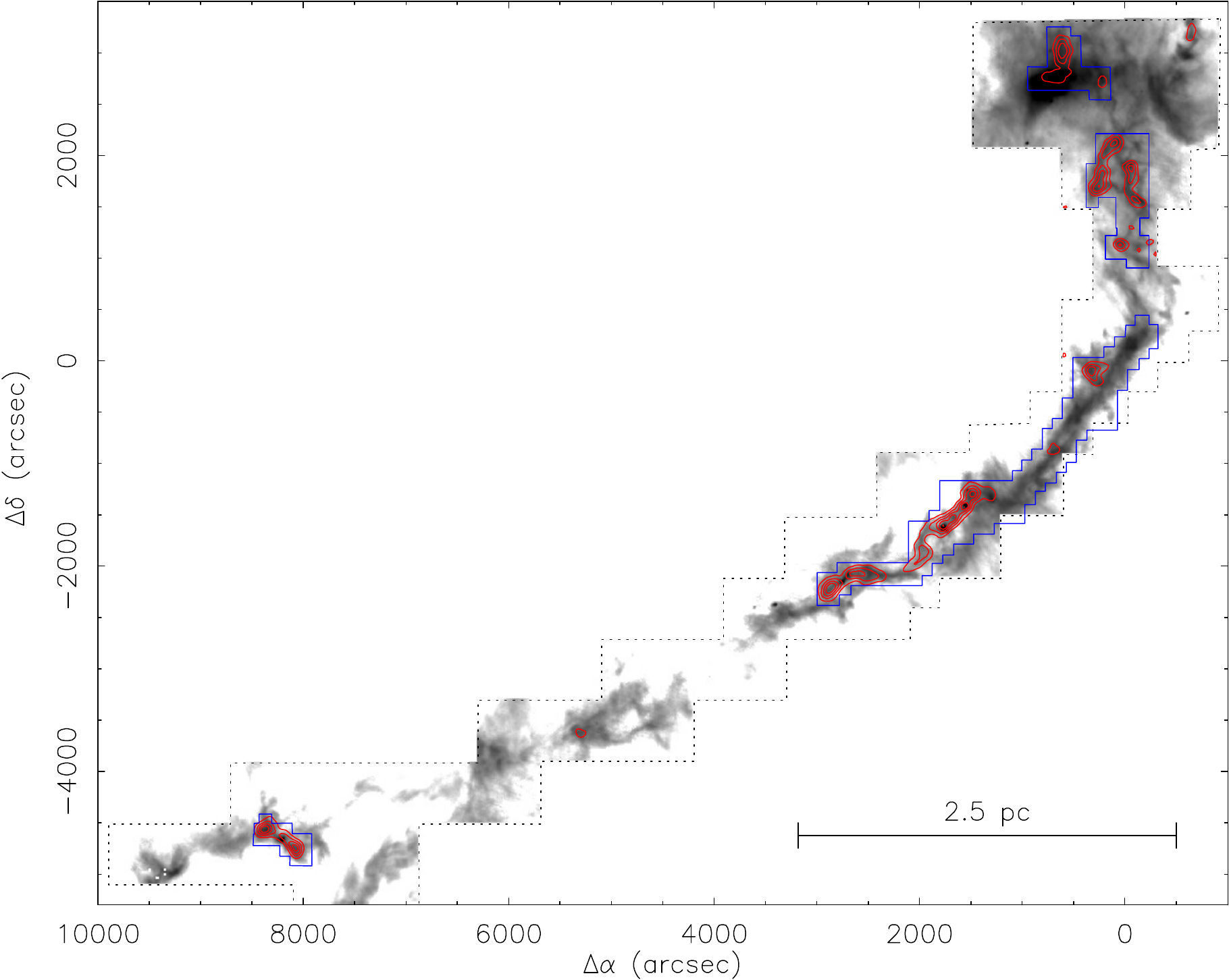}}
\caption{Large-scale view of the L1495/B213 complex.
The grey scale shows the Herschel-SPIRE
250~$\mu$m emission mapped by \citet{pal13},
and the
red contours represent the N$_2$H$^+$(1--0) emission
mapped by \citet{hac13} with  the FCRAO telescope.
The black dashed lines show the limits of the FCRAO observations,
and the blue solid lines enclose the regions newly mapped with the
IRAM 30m telescope.
\label{large_scl}}
\end{figure}

Figure~\ref{large_scl} presents a large-scale view of 
the L1495/B213 complex. The grey background represents
the Herschel-SPIRE
250~$\mu$m dust continuum emission
mapped by \citet{pal13}, which traces the 
distribution of material in the region. As can be seen, 
this emission delineates a 
$\approx 10$-pc long filament that has L1495 
at its northern end and runs almost diagonally towards 
the south-east, becoming weaker and more fragmented along
the way. This filamentary geometry seen with Herschel 
is in good
agreement with previous maps of the 
large-scale dust extinction, both
in the optical \citep{gai84,cer85}
and the NIR \citep{sch10}.

\begin{figure*}
\centering
\resizebox{\hsize}{!}{\includegraphics{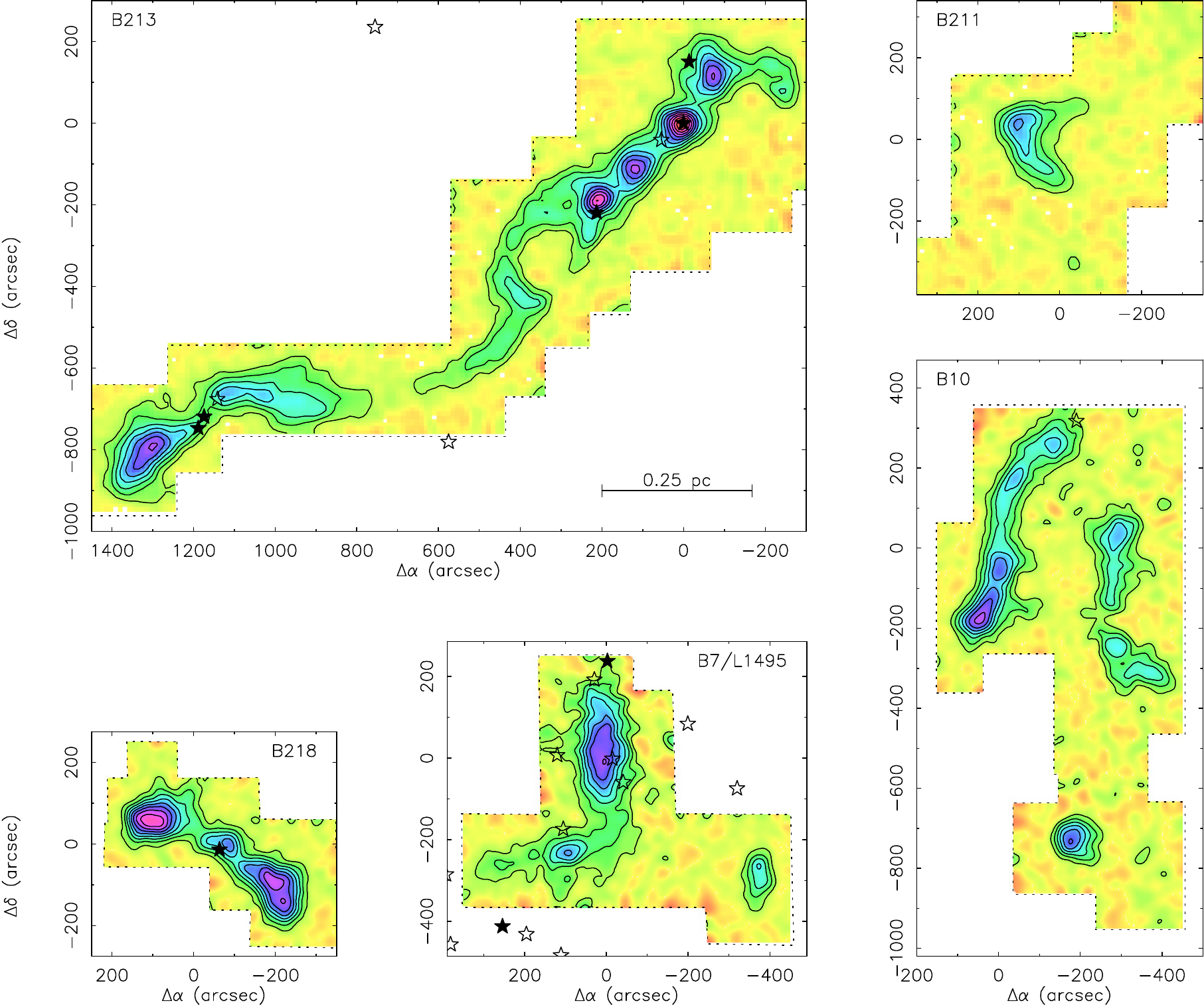}}
\caption{Integrated intensity maps of the N$_2$H$^+$(1--0) emission
showing how the dense cores in L1495/B213
form linear chain-like structures (IRAM 30m data).
All maps have the same linear and intensity scales, and the
first contour and contour interval are 0.5~K~km~s$^{-1}$.
The star symbols indicate the position of the YSOs in the
compilation of \citet{luh10}, with solid symbols
representing Class I objects.
The central positions, in ($\alpha(\mathrm{J}2000)$, $\delta(\mathrm{J}2000)$)
coordinates are
(04:19:42.5, +27:13:36) for B213,
(04:18:05.0, +27:35:16) for B211,
(04:28:02.0, +26:19:32) for B218,
(04:18:32.2, +28:27:18) for B7, and
(04:18:04.0, +28:08:14) for B10.
\label{n2hp_maps}}
\end{figure*}

While striking for its prominence, the L1495/B213 
region is not the only large-scale filamentary
structure in the Taurus dark cloud. 
Images of Taurus using both
molecules and dust reveal that most of
the material in the cloud is distributed in 
a network of crisscrossing filaments of
different sizes and orientations
\citep{bar07,dob05,lom10,gol08,kir13}.
In this sense, the L1495/B213 region is a prominent but still
representative part of the Taurus molecular cloud.

Fig.~\ref{large_scl} also shows 
the N$_2$H$^+$(1--0) emission mapped
by \citet{hac13} using the FCRAO telescope 
with an angular resolution of about $60''$
(red contours).
This N$_2$H$^+$ emission traces
the denser, chemically-evolved gas
that has condensed out of the more diffuse material
in the cloud. This gas occupies only a small fraction of
the total cloud area and forms
linear structures 
with typical length of $\approx 0.5$~pc
and generally aligned with the large-scale
direction of the cloud.
The higher resolution IRAM 30m 
observations presented here 
cover the regions indicated by
blue lines in the figure.

\subsection{High-resolution N$_2$H$^+$ maps}

Figure~\ref{n2hp_maps} shows the new IRAM 30m
maps of N$_2$H$^+$(1--0) integrated intensity
toward the regions with 
significant emission, each labeled 
with a Barnard cloud name following 
the convention of \citet{hac13}. 
The figure also shows
the location of the
young stellar objects (YSOs) identified by the Spitzer survey of
\citet{luh10} (see \citealt{reb10} for a similar compilation).

The new IRAM 30m maps agree 
with the lower resolution FCRAO maps 
of \citet{hac13}, and with maps
of some of the individual regions previously 
presented by \citet{lee01}
and \citet{tat04}.
These new maps provide a sharper and more sensitive view
of the dense cores, and highlight
the tendency of the cores to lie in elongated structures,
which we will refer to as
``chains'' due to their linear 
appearance and the presence of multiple 
peaks. 

Many N$_2$H$^+$ peaks in 
Fig.~\ref{n2hp_maps} correspond to
dense cores in the sense used by previous
studies of NH$_3$ or N$_2$H$^+$ emission, like
those of \citet{ben89} and  \citet{cas02}.
This is the case of the peaks in the B218 chain
and those in
the northern part of B213. These N$_2$H$^+$ peaks have
a strong degree of central concentration and typical
sizes of 0.05-0.1~pc, implying that the emitting gas
is self-gravitating and likely evolving
toward star formation.
Indeed, several of these peaks contain 
embedded Class 0 or Class I YSOs indicating that
star formation has already taken place.

Not
all the N$_2$H$^+$  peaks in Fig.~\ref{n2hp_maps} however
have the distinct central concentration
that we normally associate with dense cores.
Most peaks in the B10 chain, for example,
are true emission maxima, but they barely stand out 
above the extended filamentary emission that surrounds them.
This lack of contrast and central concentration makes
the nature of these peaks unclear.
They seem to represent pockets of dense gas that
are still connected to their surroundings,
and that therefore have not yet evolved into gravitationally
decoupled objects similar to the standard dense cores.
If so, these N$_2$H$^+$  peaks must correspond to an evolutionary
stage earlier than the dense core phase, but that is more evolved
than the extended material traced with C$^{18}$O by \citet{hac13}
since the gas is chemically evolved.
The variety of peak contrasts 
in Fig.~\ref{n2hp_maps} therefore implies that the 
N$_2$H$^+$ data cover
an almost complete sequence of core evolutionary stages
that goes 
from the barely-discernible 
peak near $\Delta\delta=200''$ in B10
to the highly compact peak near the origin of the B213 map,
which is associated with the well-known outflow source 
IRAS 04166+2706 \citep{bon96,taf04b,san09}.

The ambiguous classification of some of the N$_2$H$^+$ peaks
is likely a consequence of the continuous 
transition between the
ambient and core regimes, and implies that
distinguishing between ``true'' dense cores and 
precursors of dense cores is an impossible (or arbitrary) task.
For this reason, here we will not 
attempt to distinguish between 
N$_2$H$^+$ peaks that we consider to be
``cores'' and those that
we consider as not having 
reached the core phase yet,
and we will treat them equally in our analysis.
The lack of a clear distinction between
cores and pre-cores, however,
seems not enough reason to stop using the
term ``core.'' 
It is just a reminder that when studying 
core formation, some intermediate structures will
unavoidably end up having an
ambiguous nature.

\subsection{Mean surface density of companions}

\citet{hac13} used the mean surface density of companions (MSDC) to
quantify the clustering of dense cores in L1495/B213.
The MSDC measures the average number of neighbors
per unit area that one object has as a function of angular separation.
It has often been used, together
with its equivalent the two-point correlation function, 
to determine the degree of clustering of
stars in Taurus \citep{gom93,lar95,sim97,har02}.
Studying the distribution of low-density condensations 
identified from from NIR extinction measurements,
\citet{sch10} found an excess of power in the MSDC at
small angular separations. \citet{hac13} found also an 
excess of nearby companions, but in a different type
of structures: the dense cores derived from 
N$_2$H$^+$ observations.

The new IRAM 30m data provide an improved description of the
N$_2$H$^+$ emission in L1495/B213, especially at small
angular scales. Thus, as a first step in our analysis,
we have re-evaluated the MSDC of dense cores in the cloud.
To do this, we
have determined the location of all the N$_2$H$^+$ emission peaks 
in the maps of Fig.~\ref{n2hp_maps}. We have counted 
22 distinct peaks, which are three more than the 19 found
by \citealt{hac13} because the new data identify additional 
peaks in the B10 region.
Using the position of these N$_2$H$^+$ peaks,
we have calculated the MSDC following
the procedure described in \citet{hac13}.
To extend the MSDC to the largest angular separations, we have used
the large-scale information from the FCRAO data, since the new 
the new 30m observations do not recover new peaks in regions
with no FCRAO detection, but only separate better the 
regions already known to have bright N$_2$H$^+$ emission.

\begin{figure}
\resizebox{\hsize}{!}{\includegraphics{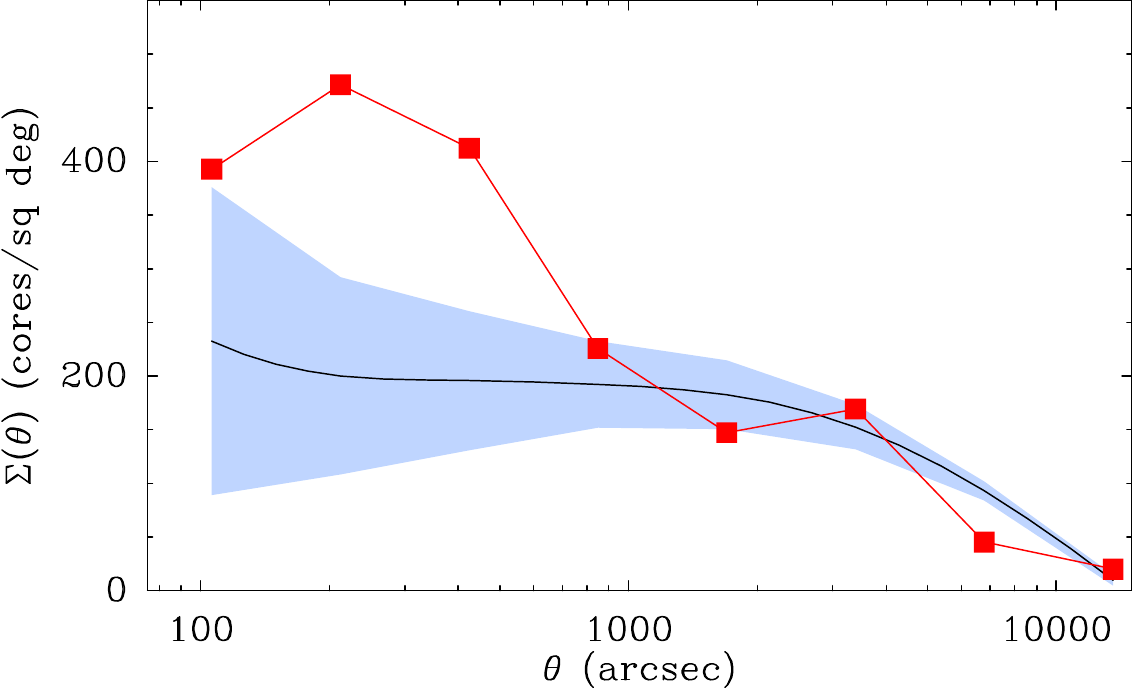}}
\caption{Mean surface density of companions 
as a function of angular separation
for the 22 cores identified in L1495/B213.
The red squares represent observed data.
The black line and the blue-shaded region represent 
the mean and the $\pm$rms interval from 100
Monte Carlo simulations of a random distribution of
22 cores.
The departure of the red
squares from the shaded region at small angles indicates
a significant level of clustering at scales less than $700''$.
\label{msdc}}
\end{figure}

The new MSDC determined from the IRAM 30m data is shown 
in Figure~\ref{msdc} with red squares.
This MSDC has a 
finer sampling than that determined with the FCRAO 
data due to the 
higher sensitivity and resolution of the new observations,
but apart from that, the two MSDCs
are consistent with each other.
Also shown in Fig.~\ref{msdc} is the
expected MSDC for a uniformly random distribution of
cores. This distribution was determined using a set
of 100 Monte Carlo simulations in which 
22 dense cores were assigned random coordinates
inside a rectangular region of dimensions approximately
equal to those of L1495/B213.
The mean value of this model MSDC is indicated by the black line, and
its rms interval is contained in 
the blue-shaded region.

As Fig.~\ref{msdc} shows, 
the MSDC for the cores in L1495/B213
has a significant excess over the random distribution for  
angular separations smaller than about $700''$
(0.5~pc for a Taurus
distance of 140~pc, \citealt{eli78}).
This excess means that for a given core, the
probability of having a neighbor
closer than about 0.5~pc
is higher than it would be if the cores were distributed 
randomly over the cloud.
The excess 
confirms the
visual impression from Fig.~\ref{large_scl} that the
N$_2$H$^+$ peaks in L1495/B213 tend to cluster
in small chains, and that the chains have a typical length 
of 0.5~pc. The chains, therefore, 
are true physical structures,
and not mere chance groupings of dense cores
in the cloud.

The clustering of dense cores into chains 
implies that core formation in L1495/B213
is a highly correlated process.
A number of authors have previously emphasized 
that core formation results from the fragmentation of 
filamentary clouds \citep{sch79,lar85,har02,mye09,and10,mol10}.
Our observations of L1495/B213 go further than that by showing that 
fragmentation does not occur equally
distributed along the length of a filament, but that it
favors special locations where multiple cores are formed in chains.
The reason for this selectivity is likely associated with the
multiplicity of filaments found by \citet{hac13},
who argued that the L1495/B213 large-scale filament is
in fact a collection of 35 intertwined velocity-coherent
filaments.
If most of these filaments are ``sterile'' and do not form
cores, while a small minority are ``fertile'' and form
multiple cores, a clustering of the cores into a few elongated
chains is naturally expected. In Sect.~4 we will
discuss in more detail the implications of this result
to our understanding of core and star formation.

\subsection{Comparison with SPIRE data and N$_2$H$^+$ abundance}

\begin{figure*}
\resizebox{\hsize}{!}{\includegraphics{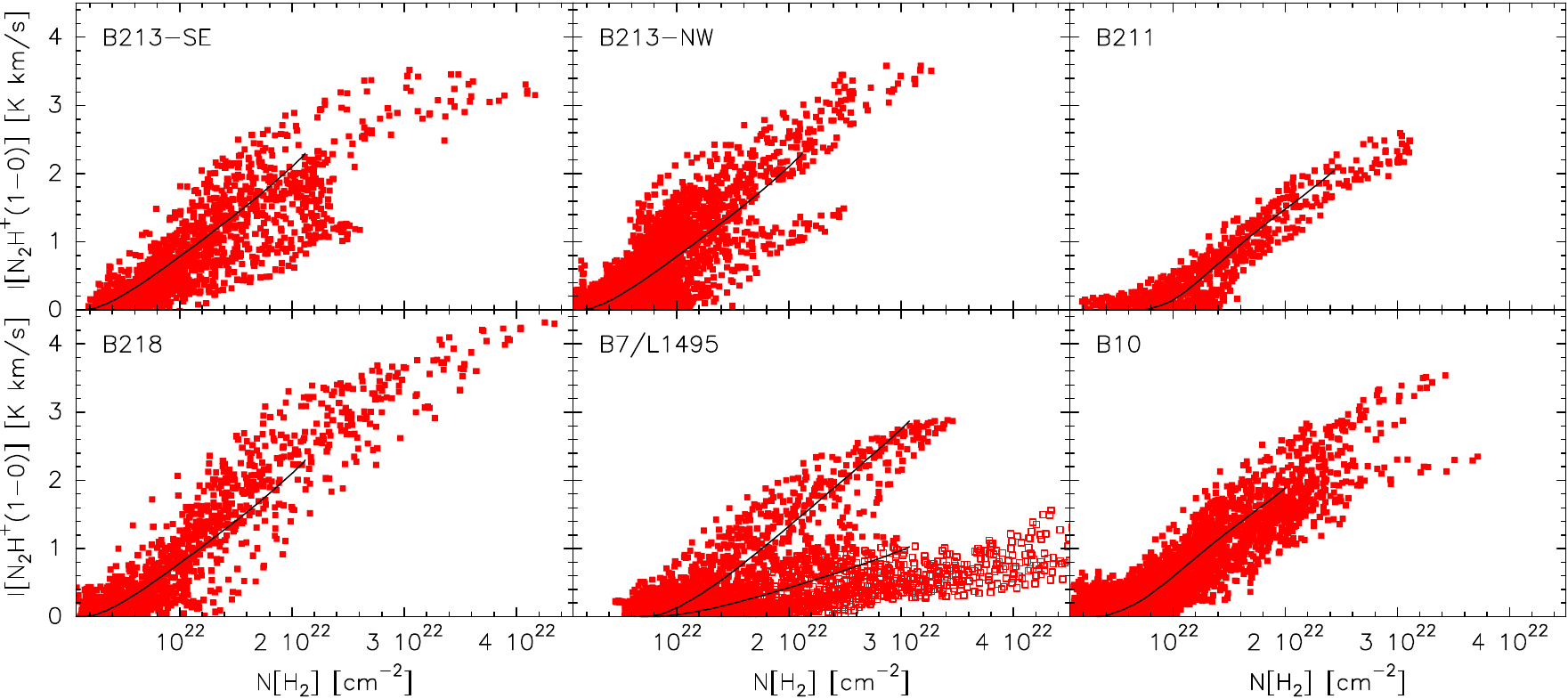}}
\caption{Correlation between observed N$_2$H$^+$(1--0) integrated intensity 
and H$_2$ column density derived from SPIRE dust-continuum data.
Points within $45''$ of an embedded YSO have been excluded
to avoid effects of stellar heating.
The open squares in the B7/L1495 panel indicate 
points likely affected by V892 Tau (see text).
The dashed lines represent the analytic expression 
discussed in the text. 
\label{dustcol}}
\end{figure*}

While
N$_2$H$^+$ is a tracer of choice for dense core gas due to its 
resistance to freeze out,
its formation is enhanced by the disappearance of CO 
from the gas phase, so it can suffer systematic
abundance variations during the evolution of a core
\citep{cas99,ber02,taf02,aik05}.
To quantify possible N$_2$H$^+$ abundance variations in the chains of
L1495/B213, we need to compare
the N$_2$H$^+$ data with data from a different tracer that is
insensitive to freeze out. As mentioned before, observations
of the dust emission and absorption provide 
an independent estimate
of the gas column density, 
and therefore
represent the ideal counterpart to the molecular
line data presented here.

The Herschel Gould Belt Survey (HGBS)
used the Herschel Space Observatory
to produce very high quality images
of the dust continuum emission from the Taurus molecular cloud
\citep{and10,pal13,kir13}.
These publicly-available images are an excellent
counterpart to the N$_2$H$^+$ line data, since they have high angular 
resolution, cover multiple wavelengths, and trace optically thin
emission.
As a fist step, we
compared the dust and the N$_2$H$^+$ data by superposing the
N$_2$H$^+$(1--0) maps of Fig.~\ref{n2hp_maps}
with the 250, 350, and 500 $\mu m$
maps made by the HGBS team with the SPIRE instrument. 
These SPIRE maps cover the longest wavelengths observable with Herschel,
and provide the highest sensitivity to cold dust emission.

The SPIRE-N$_2$H$^+$ comparison showed that in places with 
significant N$_2$H$^+$(1--0)
emission ($\ge 0.5$~K~km~s$^{-1}$), the dust continuum 
flux and the N$_2$H$^+$(1--0)
integrated intensity have similar spatial distributions.
This implies that in N$_2$H$^+$-bright places, most of the dust 
continuum emission arises from the component emitting N$_2$H$^+$
(in N$_2$H$^+$-weak places, the dust emission traces the extended cloud).
As a result, we can use the dust continuum emission from the chains to
estimate an associated  H$_2$ column density, and with
it, an N$_2$H$^+$ abundance.
To carry out this estimate, 
we first  convolved the 250 $\mu m$ map to match the 35\farcs2
angular resolution of the 500 $\mu m$ map, which is also similar to
the $33''$ resolution of the convolved N$_2$H$^+$ data (Sect.~2).
We then followed standard practice and assumed that the dust
emitted as an optically thin grey body
with an emissivity that varies with frequency as $\nu^2$ 
\citep{hil83}, and used the fluxes at the two wavelengths
to derive a dust temperature and an H$_2$ column density
for each position.

Our choice of the 500 $\mu m$ dust opacity was
$\kappa_{500 \mu\mathrm{m}}=0.03$~cm$^2$g$^{-1}$,
based on the matching between our SPIRE-derived H$_2$
column densities with the extinction-derived column densities
of \citet{sch10}, which were kindly provided by Markus Schmalzl.
This choice is only 20\% lower
than the value assumed by the HGBS team 
\citep{and10,kon10,arz11,pal13,kir13},
and lies within the range of values used or derived by other authors 
from Herschel data \citep{hen10,juv11,lau13,suu13}.
Still, it should be noted that 
the dust opacity has an uncertainty of at least 50\%, and that
it could suffer variations with 
density, as shown by the detailed analysis of \citet{juv11}, 
\citet{suu13}, and \citet{ysa13}.
The uncertainty in the dust opacity represents 
the largest source of uncertainty in our H$_2$ column density
estimate. 

Fig.~\ref{dustcol} compares the SPIRE-derived H$_2$ column density with
the
N$_2$H$^+$(1--0) integrated intensity in all the regions with 
N$_2$H$^+$ emission.
The B213 data have been split into two panels to distinguish
the region two main components (labeled SE and NW), and the southern part of 
B7/L1495 has been treated separately from the north one
(and labeled with open squares),
due to its anomalous N$_2$H$^+$ abundance further discussed below.
As the figure shows, the N$_2$H$^+$(1--0) integrated intensity
and the H$_2$ column density appear to be correlated in all regions.
An estimate of the Pearson's $r$ coefficient confirms this impression
and returns values 
that range from  0.81 in B213-NW (lowest) to 0.94 in B218 (highest),
all indicative of a significant degree of correlation.

While significant, the correlation in the panels of Fig.~\ref{dustcol}
presents a non-negligible level of scatter. Part of it 
appears to arise from the contribution of gas with either low
density or low N$_2$H$^+$ abundance (or likely both).
This gas does not contribute to the N$_2$H$^+$
intensity, but increases the H$_2$ column density and 
shifts some of the points along the x-axis, broadening
the distribution in the plots. An extreme example of this effect 
can be seen in the B213-NW panel, where in addition to the main
diagonal band of points there is a secondary band that has
weaker N$_2$H$^+$ emission and is shifted horizontally by about 
$10^{22}$~cm$^{-3}$. This band
is associated with a small condensation near $\Delta\alpha=-200''$
seen in Fig.\ref{n2hp_maps}.
A less extreme example occurs in B218,
where the broad diagonal band of points is in fact the superposition of 
two slightly shifted and
narrower bands, each one due to one of the
bright cores in the chain.
In contrast, the B211 region presents only a single dense
core in the N$_2$H$^+$  maps, and its distribution of points 
presents the narrowest correlation
of the sample. 

If the scatter in the panels of Fig.~\ref{dustcol}
results from additional gas components along the 
line of sight, the slope 
is an indicator of the N$_2$H$^+$ abundance in the dense gas. This is
expected because the N$_2$H$^+$ integrated intensity represented 
in the y-axis is proportional to the N$_2$H$^+$
column density (assuming that the emission is optically
thin, see below), and as a result, the slope of the correlation
equals the ratio between the N$_2$H$^+$ and H$_2$ column 
densities, which is an estimate of the N$_2$H$^+$ abundance. 
A detailed radiative transfer model
presented in the next section to fit
the radial profiles of N$_2$H$^+$  emission
confirms this interpretation, and shows that
there is an almost linear relation between the 
N$_2$H$^+$ integrated intensity
and the H$_2$ column density. 
This is illustrated in Fig.~\ref{dustcol} with a series of     
black lines that represent the predictions from the radiative
transfer model
assuming that all chains have the same N$_2$H$^+$
abundance of
$5 \times 10^{-10}$ with the exception of 
B7/L1495-south, where the abundance 
is $1.5 \times 10^{-10}$.
To fit the data in Fig.~\ref{dustcol}, we added
small horizontal offsets of $7 \times 10^{21}$~cm$^{-3}$, 
$6 \times 10^{21}$~cm$^{-3}$, and $3 \times 10^{21}$~cm$^{-3}$ 
to the fits of B211, B7/L1495-north, and B10,
respectively.
The offset in B211 is in
fact expected, since this region contains 
two additional C$^{18}$O components (numbers 9 and 12 in the 
decomposition of \citealt{hac13}) that do not emit in N$_2$H$^+$
and clearly contribute to the H$_2$ column density
(see Fig.~\ref{c18o_2vel} below).
The offsets
in B7/L1495 and B10 are also likely related to the presence of
lower-density gas toward these two regions.

While uncertain, the N$_2$H$^+$ abundance in
B7/L1495-south is much lower than 
in other chains (by a factor of 3), and is the only
one that deviates from a pattern of almost constant
abundance.
To investigate its origin, 
we have inspected the SPIRE images of B7/L1495
at different wavelengths.
These images show that B7/L1495-south lies inside
a region of
bright and extended FIR emission in the vicinity of V892 Tau, a
Herbig Ae/Be star first identified by \citet{eli78}
and with a total luminosity of
$\approx 400$~L$_\odot$
\citep{san11,moo13} that lies about $300''$ (0.2~pc) 
in projection from B7/L1495-south.
A grey body analysis of the SPIRE emission, 
indicates that the dust temperature in the region 
is elevated, and that it
gradually increases toward 
V892 Tau, where it reaches about 15~K, or 50\%
higher than in B7/L1495-north.
In addition, the C$^{18}$O maps of
Fig.~\ref{c18o_2vel} (discussed below)
show that B7/L1495-south coincides
with a region of bright C$^{18}$O emission. This is in
contrast with the other N$_2$H$^+$-bright regions,
which coincide with weak C$^{18}$O emission due to
CO freeze out. Thus, it appears that molecular depletion,
and its resulting N$_2$H$^+$ enhancement, are anomalously low
in the dense gas of B7/L1495-south.
The low N$_2$H$^+$ abundance in B7/L1495-south seems therefore
a result from the action of V892 Tau.
Whether this is a consequence of simple dust heating or of a
different energetic process requires a more detailed investigation.

\subsection{Density structure of the chains}
\label{sect_radprof}

\begin{figure*}
\resizebox{\hsize}{!}{\includegraphics{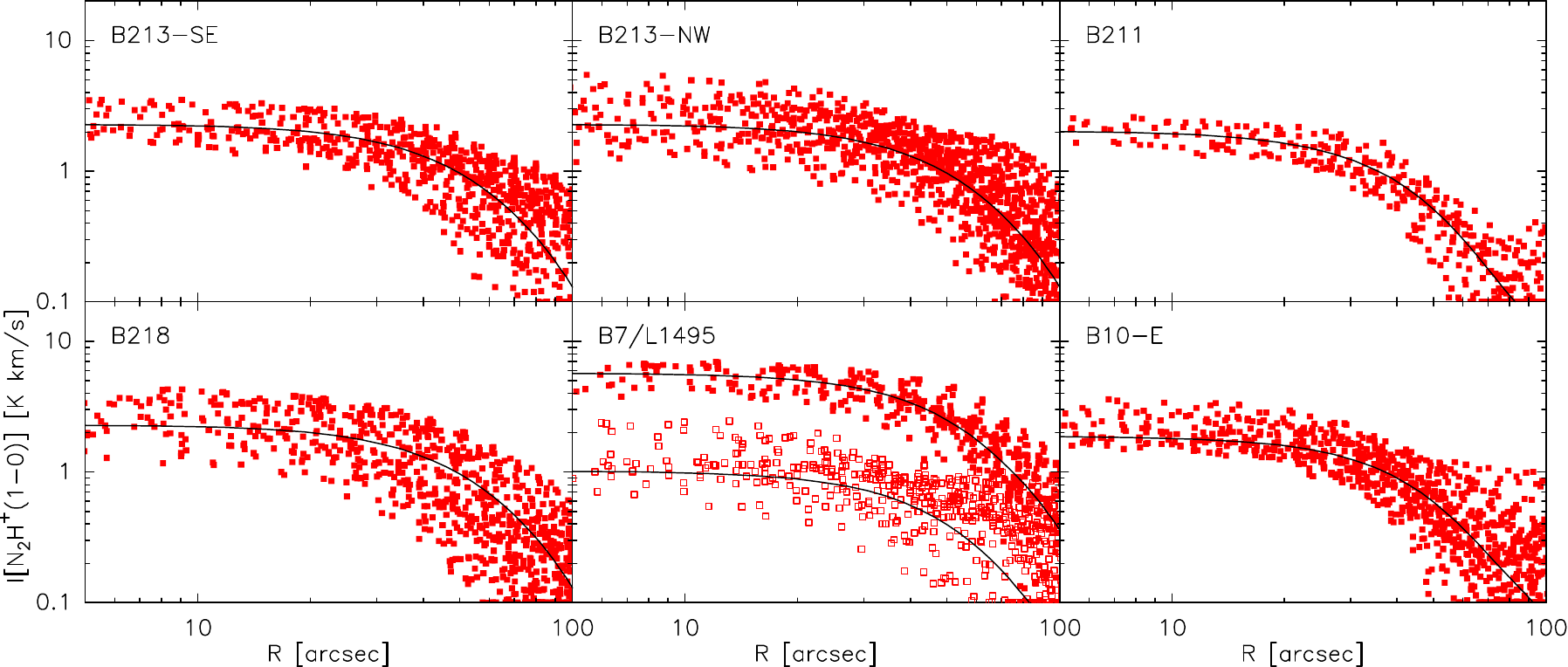}}
\caption{Radial profiles of N$_2$H$^+$(1--0) integrated intensity in
log-log scale.
The red squares represent the data, and the
black lines are our models.
B7/L1495 north data and model have been shifted by a factor of
two to ease visibility.
\label{rad_profs}}
\end{figure*}

In the maps, the core chains appear irregular in shape and
different from each other. A
closer inspection of the emission, however, shows
that they have a similar internal structure.
This can be seen in the radial profiles of 
N$_2$H$^+$ emission presented
in Fig.~\ref{rad_profs}. These profiles were created
by following
the emission of each chain in the map with a cursor
and defining the line of relative maxima as
the axis of the chain. Using this
axis, the radial distance of each observed position
was calculated, and the intensity of the emission 
was plotted as a function of it.

As figure Fig.~\ref{rad_profs} shows, the 
emission from each chain follows a 
radial profile that consists of
a flat inner region and a power-law tail,
similar to that often found in
filamentary clouds \citep{arz11,hac11,pal13}.
The B213 and B218 chains present a larger level of dispersion
near the axis 
because they contain bright cores separated by
regions of weak emission, so points with the same
axial radius can have a large range of intensities.
The B10 and B7/L1495-north chains, on the other hand,
present a less clumpy and more pristine appearance, and
their radial profiles have a lower dispersion
near the axis 
(as in Fig.~\ref{dustcol}, the emission from
B7/L1495 has been separated into north and south components).

The combination of clumpiness due to the
embedded cores and comparable radial profiles 
implies that the chains started their evolution 
with a similar density
structure, and that later events added
different fragmentation patterns to each one.
Our goal in this section is to
determine this common underlying structure, since 
it represents the initial conditions of core formation. 
For this, we have modeled the N$_2$H$^+$ radial profiles 
assuming that the chains
are cylindrically symmetric, and that they have 
a density profile of the form
\begin{equation}
\label{threepar_cyl}
n(r)=\frac{n_\circ}{1+(r/r_0)\,^\alpha},
\end{equation}
where $n_\circ$ is the central density, $r_0$ the half-density
radius, and $\alpha$ the asymptotic power index. 
This type of radial profile has previously 
been used to fit the density structure of
both starless cores \citep{taf02} and 
filamentary clouds \citep{arz11,hac11,pal13,ysa13}.

\begin{figure}
\resizebox{\hsize}{!}{\includegraphics{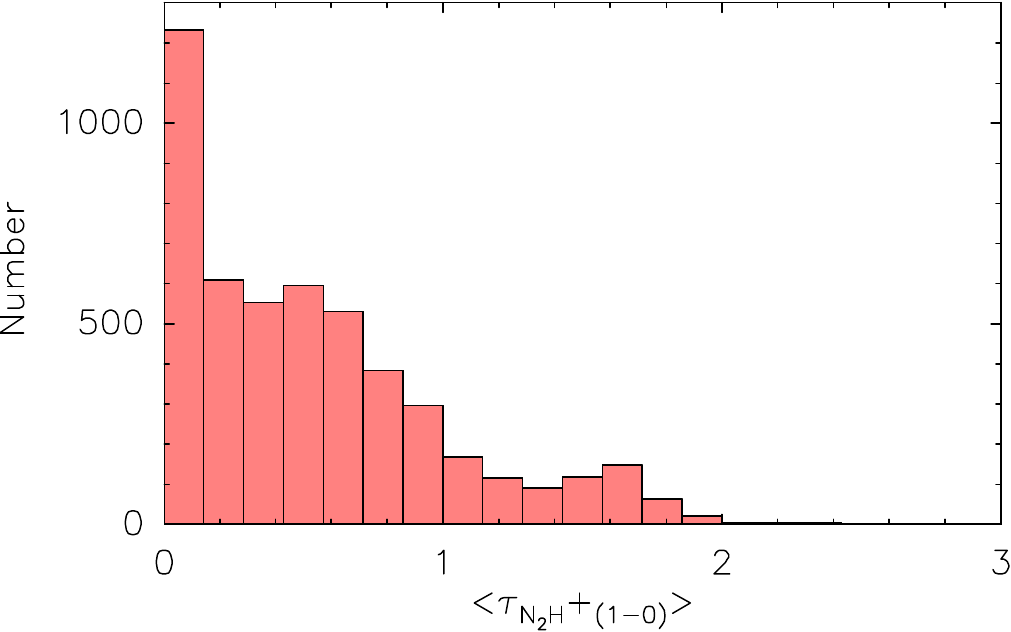}}
\caption{Histogram of the N$_2$H$^+$(1--0) mean optical depth 
for all chain positions used in the radiative transfer analysis.
\label{tau}}
\end{figure}

To compare the model 
with the observations, we have solved the equations of 
radiative transfer and predicted the radial profile of 
N$_2$H$^+$ intensity. Lacking a 2D model for the radiation
transfer in a cylinder, we have instead used a spherically
symmetric model that has the same density radial profile.
This model properly 
accounts for the radial drop of collisional excitation 
caused by the density law, although it likely underestimates
the excitation due to photon trapping because photons 
escape more easily from a sphere that from a cylinder.
The effect of this difference in the trapping, however, 
is likely to be very small, since the
optical depth of the N$_2$H$^+$(1--0) emission is low.
This is illustrated in Fig.~\ref{tau} with a histogram
of the mean N$_2$H$^+$(1--0) optical depth estimated
from the data of all bright positions in our survey 
(4,947 spectra in total).
The mean optical depth was defined by dividing 
the total optical depth of the N$_2$H$^+$(1--0) transition
(determined with the HFS hyperfine analysis 
in the CLASS program) between seven, which is the number of components.
It therefore represents the average optical
depth of an individual N$_2$H$^+$(1--0) component.

As can be seen, 
the histogram of mean optical depths
is dominated by low values.
In all regions but B7/L1495, $\sim 90$\% of the points have a mean
optical depth lower than 1, and less than 1\% of the points have
a mean optical depth larger than 2. The cluster-forming B7/L1495
region
is significantly more opaque and is responsible for the small
group of points between optical depths 1 and 2 in the histogram.
Still, 56\% of its positions have a mean optical
depth lower than 1, and no point exceeds a value of 2.
Under these conditions, using spherical geometry to 
simulate the radiative excitation in a cylinder appears to be 
an acceptable approximation, especially considering that the
assumption of cylindrical symmetry is itself a 
large simplification of the true geometry of the chains.

To solve the radiative transfer equations we used
the Monte Carlo code of \citet{ber79} 
previously applied to analyze the emission from starless cores
\citep{taf02,taf04a}. This code was implemented with 
the molecular parameters of N$_2$H$^+$ from the
LAMDA web site \citep{sch05}, which include
the collision rates of \citet{dan05}. These rates
include the individual hyperfine components
of each rotational transition (up to J=6), 
which were treated 
as independent lines whose flux was
later combined
to simulate the observed integrated 
intensity.

Our radiative transfer calculation also 
assumed a constant gas kinetic temperature of 10~K.
This assumption is
based on the analysis of ammonia,
a molecule that
coexists with N$_2$H$^+$ in the dense gas 
and whose emission in Taurus cores indicates a median temperature
of 9.5~K (which little dispersion, \citealt{jij99}). 
Recent large-scale ammonia mapping of the L1495/B213
complex by Seo et al. (in preparation) confirms this assumption,
and indicates that while there are small local variations of 1-2~K
in some cores, there are no global temperature gradients in the chains.
Even the lower-density gas that surrounds
the core chains appears to have a similar temperature,
since the CO-based estimate of \citet{gol08} indicates 
that the majority of points in this region
(their Mask 2) have temperatures that lie in the 6-12~K range.
This constant temperature of the gas in the 
density range of interest ($10^5$-$10^4$~cm$^{-3}$, see below)
is expected from detailed modeling of the gas heating and cooling, 
and contrasts with the well-measured temperature 
gradient of the dust component found by \citet{pal13}, which 
is expected from heating by the interstellar radiation field
\citep{eva01,gal02}.

Additional assumptions of the model were a non-thermal 
FWHM linewidth of 0.25~km~s$^{-1}$, as suggested by 
the
analysis of Sect.~\ref{sect_nonth}, and a maximum 
radius of 400~arcsec (0.27~pc), although the comparison
with the data is restricted to the central 100~arcsec
due to limited signal to noise (the exact size
has only a small effect on the result). Following our experience
with the analysis of dense cores, we divided the cloud
model into 200 shells, used 2,000 photons, and iterated
the calculation 40 times. To simulate the IRAM 30m
observations, plus the additional $20''$ Gaussian
smoothing applied to the data
to enhance its signal to noise, the emerging intensity
distribution was convolved with a Gaussian 
of $33''$ FWHM.
Finally, the data were scaled up by a factor of 
1.4 to simulate a 45 degree inclination angle of 
the model with respect to the line of sight.
The L1495/B213 cloud appears as a
relatively long filament in the sky ($\approx 10$~pc),
so it is unlikely to be highly inclined; using a
moderate angle of 45 seems like
a reasonable assumption that is unlikely to introduce
a large error.

\begin{table}
\caption[]{Best-fit chain parameters.\label{tbl_fit}}
\centering
\begin{tabular}{lcccc}
\hline
\noalign{\smallskip}
Chain  &   $n_0$        &      $r_0$      &     X(N$_2$H$^+$) &  $M/L$ \\
 &   (cm$^{-3}$)        &      $('')$      &     &  ($M_\circ$~pc$^{-1}$) \\
\noalign{\smallskip}
\hline
\noalign{\smallskip}
B213-SE & $6 \times 10^4$ & 50 &  $5 \times 10^{-10}$ & 33 \\
B213-NW & $6 \times 10^4$ &  50 &  $5 \times 10^{-10}$ & 33 \\
B211 & $7 \times 10^4$ &  35 &  $5 \times 10^{-10}$ & 19 \\
B218 & $6 \times 10^4$ & 50 &  $5 \times 10^{-10}$ & 33 \\
B7/L1495-N & $7 \times 10^4$ & 50 &  $5 \times 10^{-10}$ & 38 \\
B7/L1495-S & $7 \times 10^4$ & 50 &  $1.5 \times 10^{-10}$ & 38 \\
B10-E & $6 \times 10^4$ & 40 &  $5 \times 10^{-10}$ & 33 \\
B213-SE & $6 \times 10^4$ & 50 &  $5 \times 10^{-10}$ & 21 \\
\hline
\end{tabular}
\end{table}

To find the best fit to the data, we explored different
values of the N$_2$H$^+$ abundance and 
the density law. As discussed in the previous section, the choice
of N$_2$H$^+$ abundance has a direct effect on the slope of 
correlation between 
the N$_2$H$^+$ intensity and the H$_2$ column density,
and the plots of Fig.~\ref{dustcol} were used to derive
a constant value of $5\times 10^{-10}$ for all chains but
B7/L1495-south (were the best fit is $1.5\times 10^{-10}$).
To fit the density law, we used the radial profiles of
N$_2$H$^+$(1--0) intensity and explored the effect of each of the
three free parameters. The power law index is only weakly
constrained, since the radial profile only approaches
this asymptotic behavior at large radius, where the data have
a low signal to noise ratio. 
Reasonable fits were achieved with values
close to 3 (as found for the filaments in L1517, \citealt{hac11}),
so this parameter was fixed to 3 in all the chains.
The remaining two parameters, central density and half-maximum
radius, are somewhat correlated, since both contribute
linearly to the central column density, and they need to
be distinguished by fitting the profiles at large radii 
(with the already mentioned problem of low signal to noise 
and certain dependence on the power-law index). 
After exploring a number of combinations, we
determined as best fit values those given in Table~\ref{tbl_fit},
which produce the radial profiles shown with black lines in
Fig.~\ref{rad_profs}.
Since we were interested in the density structure of
the chains as possible indicator of the initial conditions of 
core formation, the fits were
purposely chosen to fit the points with lowest intensity
near the axis and to avoid the
brighter points that arise from the dense cores.

As can be seen in Table~\ref{tbl_fit}, both the 
central density and the half-maximum radius vary
little over the sample of chains 
($6-7 \times 10^4$~cm$^{-3}$ and $35''-50''$, respectively).
This small variation agrees with our expectation of a common
internal density structure based on the similarity of the
radial profiles, and strengthens the idea that
the different chains may have formed in a similar manner. 
The fit values, however, have a significant
level of uncertainty due to the uncertainty in the
the dust opacity discussed before. Also, the 
large scatter in the radial profiles
is a remainder that cylindrical symmetry
is an over-simplification of the 3D geometry of
the chains. For this reason, the best-fit parameters 
in Table~\ref{tbl_fit} should be considered only as a first-order
approximation to the true parameters of the chain gas,
which likely have an uncertainty level of a factor of 2.

Even if approximate, the parameters of Table~\ref{tbl_fit} can be used
to explore the physical state and gravitational 
stability of the core chains.
To do this, we compare our best fit models with the classical
solution of an isothermal cylinder in equilibrium, first
studied by \citet{sto63} and  \citet{ost64}. 
This solution has an asymptotic power-law index of -4,
while our best fit models are slightly flatter and have
a power-law index of -3. More importantly, the isothermal
cylinder has an equilibrium mass per unit length of 
16.6~M$_\odot$~pc$^{-1}$, assuming a gas kinetic
temperature of 10~K. Table~\ref{tbl_fit} shows that the
mass per unit length values of our best-fit models are
systematically larger, although only by at most a factor of 2. 
\citet{pal13} also found a larger-than-equilibrium mass per unit
length in the B213/B211 filament as a whole using dust continuum
measurements, although these authors treated the region as a single
object and ignored the presence of multiple velocity components.
Whether this larger mass per unit length means that 
the chains are significantly out of equilibrium
is unclear, especially considering the uncertainty in the
dust opacity and that additional support mechanisms, such as
magnetic fields or temperature gradients can increase
the equilibrium mass per unit length \citep{sto63,nak93,rec13}.
Further understanding of the physical state of the chains requires
the analysis of their internal kinematics, which is the topic of the
next two sections.

\subsection{Cloud kinematics from C$^{18}$O data: multiple components}
\label{sect_c18o}

\begin{figure*}
\resizebox{\hsize}{!}{\includegraphics{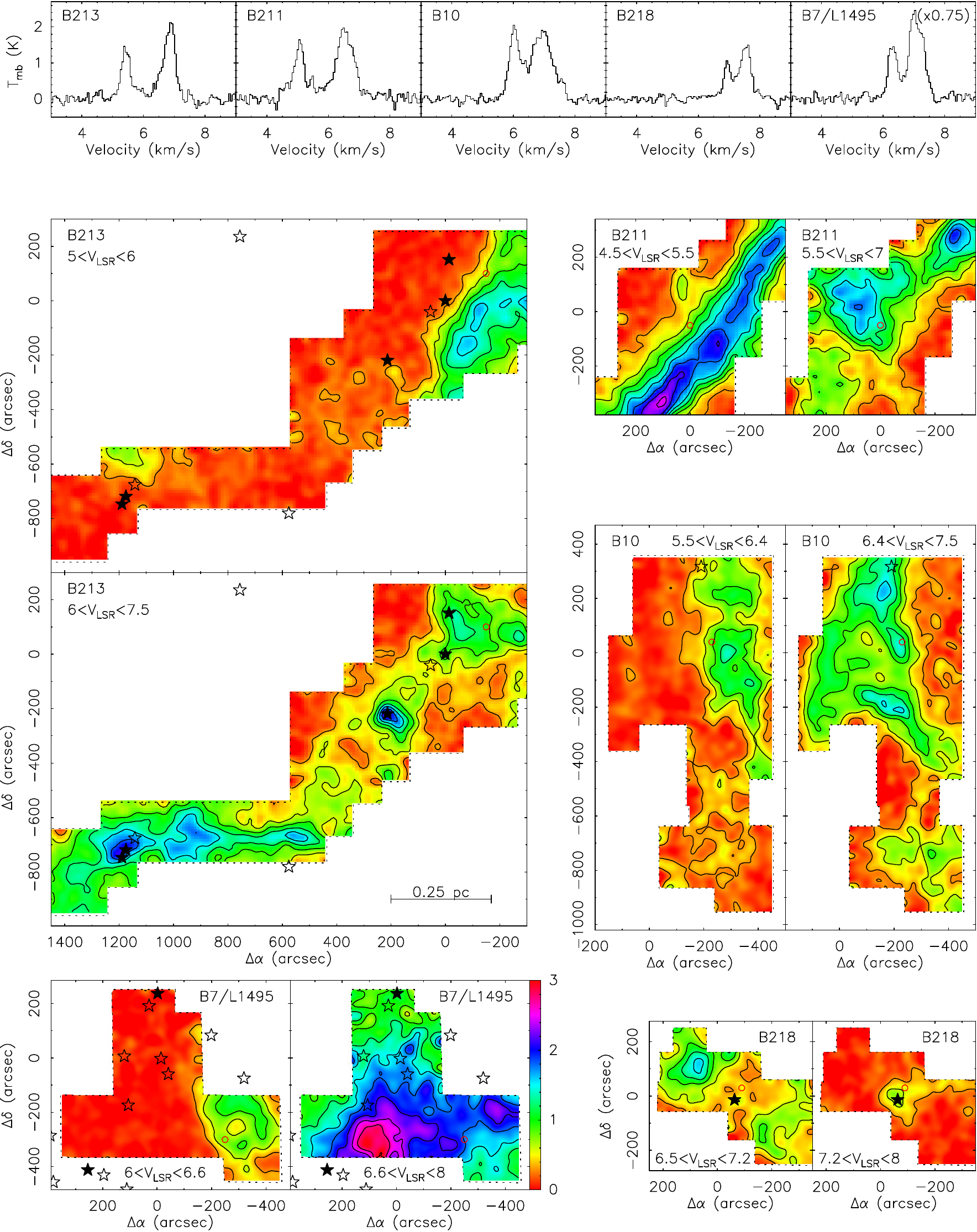}}
\caption{Velocity structure of the C$^{18}$O(2--1) emission
in  L1495/B213.
{\em Top:\ } C$^{18}$O(2--1) spectra from selected positions
illustrating the presence of multiple velocity
components along the line of sight. The selected positions are 
indicated with red circles in the maps below.
{\em Bottom:\ } Maps of C$^{18}$O(2--1) emission integrated in two
velocity intervals that approximately coincide with the components
in the top spectra.
All maps have the same physical scale, color code (shown in the B7/L1495
panel), and contour scale (first contour and interval are 0.3~K~km$^{-1}$).
Coordinate centers and star symbols as in Fig.~\ref{n2hp_maps}.
\label{c18o_2vel}}
\end{figure*}

The C$^{18}$O molecule freezes out rapidly onto the dust grains at 
densities typical of the cores and the chains, so it is 
a poor tracer of the dense gas kinematics. 
It is however a faithful tracer of the motions in the lower-density 
gas that surrounds the chains, since in this regime C$^{18}$O
is chemically stable, 
easily thermalized, and does not suffer appreciably from saturation
due to its low abundance. \citet{hac13} showed
that in the L1495/B213 region, the
velocity fields of C$^{18}$O and N$_2$H$^+$ are similar, 
indicating that the dense cores and their surrounding 
environment are closely coupled kinematically.
Before studying the kinematics of the dense gas with N$_2$H$^+$
in the next section, it is
therefore convenient to use the C$^{18}$O emission to
determine the properties of the velocity 
field in the vicinity of the chains. These properties 
provide important context and help solve some
of the ambiguities that affect the more selective
N$_2$H$^+$ emission. 

As analyzed in detail by \citet{hac13}, the 
C$^{18}$O velocity field in L1495/B213 is complex. It 
consists of about 35 intertwined filamentary components
(or fibers)
that appear in the spectra as multiple velocity peaks.
To disentangle these components, \citet{hac13} used
a combination of Gaussian fits to the spectra and the 
Friends In VElocity (FIVE) algorithm, which connects spatially 
the velocity
components from nearby positions.
The IRAM 30m maps discussed here are less extended than the
FCRAO maps of \citet{hac13},
and the focus of our analysis is limited to
the C$^{18}$O emission related to the 
dense gas in the chains. For this reason, we
have carried out a simplified analysis 
of the velocity structure of the C$^{18}$O emission
based on the inspection of the
spectra and the use of velocity-integrated maps.

An inspection of the C$^{18}$O data reveals that 
each mapped region contains at least several positions
where the spectrum has two peaks separated by more than one full linewidth.
These double-peaked spectra do not originate from self-absorption,
since, when detected, the optically thin
isolated component of N$_2$H$^+$(1--0) matches the velocity of one
of the two C$^{18}$O components,
instead of appearing at the intermediate velocity that would be
expected in the case of self-absorption.
The double peaks therefore arise from the multiple velocity
components studied by \citet{hac13} when they 
overlap along some lines of sight.
Examples of these double-peaked spectra can be seen
in the top panel of Fig.~\ref{c18o_2vel}
for each of the five regions associated with dense gas.

To determine the spatial distribution of the C$^{18}$O velocity 
components in the vicinity of the core chains, we have divided the 
emission into two velocity intervals centered approximately 
on each of the C$^{18}$O peaks.
The resulting maps, presented in the
bottom panels of Figure~\ref{c18o_2vel}, 
show that in each chain, the two C$^{18}$O
components differ markedly in spatial distribution.
In B213, the blue component
extends to the NW of the mapped region and is unrelated 
to the chain of N$_2$H$^+$ cores, that has
a different velocity and spatial distribution.
This  blue C$^{18}$O component corresponds to component number 18 in the 
cloud decomposition of Hacar et al. (2013). 
The red component, on the other hand,
is associated with the chain of dense cores, and its
velocity and large-scale orientation are similar to those
of the chain.
In contrast with N$_2$H$^+$, the
C$^{18}$O emission presents strong evidence of
large-scale freeze out. It misses some of
the brightest N$_2$H$^+$ and continuum peaks, 
such as the core around
IRAS~04166+2706 and the starless core SE of it,
and only presents bright emission toward some embedded YSOs.
C$^{18}$O freeze out in this region seems therefore not limited to
the dense cores, but occurs at the
scale of the full chain, and is only reversed locally in
the vicinity of some YSOs.

In the single-core region B211, also the red component is 
associated with the N$_2$H$^+$
dense core, since it has the same velocity and a similar spatial
distribution. The unrelated blue component
arises from a long diagonal filament that 
is in fact the superposition of the parallel components 
9 and 12 in the velocity decomposition of \citet{hac13}.

The more complex
B10 region appears in the N$_2$H$^+$ maps of Fig.~\ref{n2hp_maps}
as consisting of
two roughly parallel chains plus an isolated core in the south.
The C$^{18}$O spectrum and
maps in Fig.~\ref{c18o_2vel} show now that the
western chain of B10 coincides with a region where two velocity
components that are separated by 1~km~s$^{-1}$ spatially overlap.
These two components seem to be responsible for the apparent
velocity jump of about 1~km~s$^{-1}$ seen
in N$_2$H$^+$ (Sect.~\ref{sec_linecenter}), indicating that the 
western chain of B10 n
is in fact the overlap of two separate structures.
This interpretation  
is in agreement with the
decomposition of the large-scale C$^{18}$O
emission by \citet{hac13}, who
divided this region into two components labeled 6 and 8.

The C$^{18}$O maps of Fig.~\ref{c18o_2vel} also show that 
chain-wide CO freeze out has also taken place
in B10, since the eastern chain is very
prominent in N$_2$H$^+$ but
only marginally visible in C$^{18}$O.
As mentioned before, this eastern chain shows
little fragmentation. This indicates that CO
depletion precedes the fragmentation of the chain into 
cores.

In the B7/L1495 region, the blue C$^{18}$O velocity component is
associated with the isolated N$_2$H$^+$ core to the SW, while the
red component is associated with the N$_2$H$^+$ chain of cores.
As can be seen in Fig.~\ref{c18o_2vel}, the blue C$^{18}$O brightens
significantly toward the south end of the map, which coincides with
the region where the dust temperature increases due
to heating by V892 Tau. 
This brightening of the C$^{18}$O emission implies that CO depletion
may be lower closer to V892 Tau, and this may 
explain the anomalously low N$_2$H$^+$ abundance inferred 
from the comparison with the SPIRE data. 

Finally, in B218, the two C$^{18}$O components present
anticorrelated spatial distributions.
The blue component
peaks toward the NE and SW of IRAS~04248+2612,
and is associated with the two N$_2$H$^+$ starless cores 
at each side of the YSO because they match both in
position and velocity.
The red C$^{18}$O component,
on the other hand, peaks toward the IRAS source and extends slightly toward
the NW. The nature of this component is unclear due to its
limited extent. A likely possibility is that it
is caused by the action of the YSO on its surrounding gas, since
IRAS~04248+2612 powers a molecular outflow that has a dominant
red wing (although  mostly towards the SE, see \citealt{nar12}),
and is associated with a chain of HH objects
directed opposite to the red C$^{18}$O emission \citep{gom97}.
In this interpretation, the B218 region would
consists of a single C$^{18}$O velocity component that
corresponds to the blue C$^{18}$O regime. Further observations
of this region are needed to clarify its kinematics.

To summarize, the C$^{18}$O data show that the presence 
of multiple
velocity components in the vicinity of the chains
is common. 
These components are separated by 
supersonic speeds and do not seem to be
interacting, since they are forming
dense cores at their own systemic velocity, and not 
at the intermediate velocity that would be expected 
if core formation occurred through collisions.
The components in each region, however, are not completely unrelated. 
In B10 and B7/L1495, for example, the two components have 
produced cores
in close proximity, which given the strong clustering of 
cores in the cloud seems an unlikely random
event. In other regions, like B213 and B211,
different filamentary components are almost parallel,
suggesting that they have some type of
relation or common origin. From their large-scale
study of the C$^{18}$O emission, \citet{hac13} found that 
indeed, most velocity components in L1495/B213 belong to groups
or bundles with a common origin, 
and proposed that some type of turbulent fragmentation 
process
was responsible for their origin. 
A number of recent 
hydrodynamical simulations have shown 
that bundles of filamentary components
like those in L1495/B213
arise naturally from the combination of
turbulent motions and
self gravity \citep{kri13,smi13,moe14,mye14}.
This implies that the formation of dense cores and
chains is preceded by a step of fragmentation whose
product are the C$^{18}$O components shown in 
Fig.~\ref{c18o_2vel}.
Thus, the multiplicity of components in
the C$^{18}$O spectra near the core chains is not 
a mere superposition coincidence, but
a natural consequence of 
the hierarchical fragmentation required to form dense cores.
Further discussion on this topic is presented below
after the analysis of the N$_2$H$^+$ kinematics.

\subsection{Chain kinematics from N$_2$H$^+$ data}
\label{sec_n2hpkin}

\begin{figure*}
\resizebox{\hsize}{!}{\includegraphics{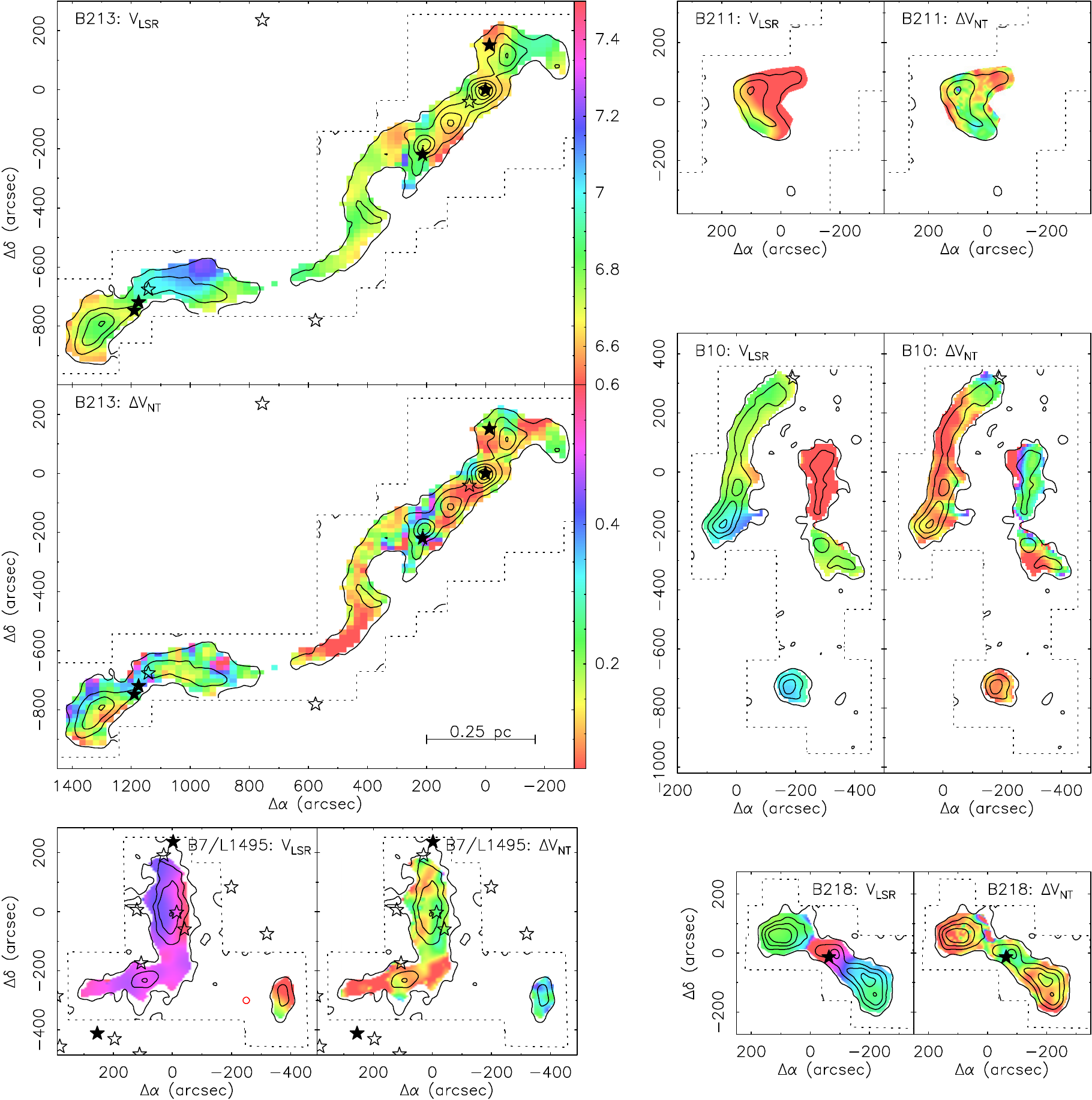}}
\caption{Velocity structure of the N$_2$H$^+$ emission 
as determined from hyperfine fits to the spectra.
For each region, the first panel shows (in km~s$^{-1}$)
the spatial distribution of the velocity centroid, and
the second panel shows (also in km~s$^{-1}$) the distribution 
of non-thermal 
linewidth (FWHM). The black contours show the distribution
of integrated emission to help identify the main dense gas features.
All plots use the same spatial scale and color code, which is indicated 
by wedges in the maps of B213.
Coordinate centers, contour levels, 
and star symbols as in Fig.~\ref{n2hp_maps}.
\label{n2hp_vel_maps}}
\end{figure*}

The velocity structure of the N$_2$H$^+$ emission is simpler than that
of C$^{18}$O due to the more selective nature of this tracer.
In general, the N$_2$H$^+$(1--0) spectra present a single
velocity component, although split into seven features due to
hyperfine structure. A few spectra show hints of 
two velocity components, like near B213 ($900''$, $-650''$),
but the components are so weak that is not possible to 
analyze them using multiple fits.
For this reason, we have fitted the 
N$_2$H$^+$(1--0) spectra assuming a single velocity component,
using for this the CLASS program 
and the numerical parameters of the 
hyperfine structure derived by  \citet{cas95}.
This single-component fit analysis determines
both the line center
velocity and the full width at half maximum
(FWHM) corrected for optical depth broadening.
Subtracting the thermal contribution of a gas at 10~K, the
FWHM can be converted into an estimate of the
non-thermal velocity dispersion in the gas.

Figure~\ref{n2hp_vel_maps} shows in color the distribution of 
N$_2$H$^+$ line center velocity 
and non-thermal FWHM as derived from the hyperfine analysis.
To ensure the quality of the data, the figure only presents results 
from fits that appear reliable 
under visual inspection, which approximately corresponds 
to an intensity threshold of 0.5~K~km~s$^{-1}$.
As can be seen, the line center velocity 
changes smoothly over each chain, with a typical
size scale for the changes of the order of a core diameter.
The accompanying linewidth maps also show a smooth behavior,
although there are several regions of high dispersion
that we discuss in more detail below.
Since the sound speed linewidth corresponds to
0.45~km~s$^{-1}$, the maps in Fig.~\ref{n2hp_vel_maps}
indicate that the gas in the chains is
mostly subsonic, and that only a few locations have 
supersonic linewidths.

\begin{figure*}
\resizebox{\hsize}{!}{\includegraphics{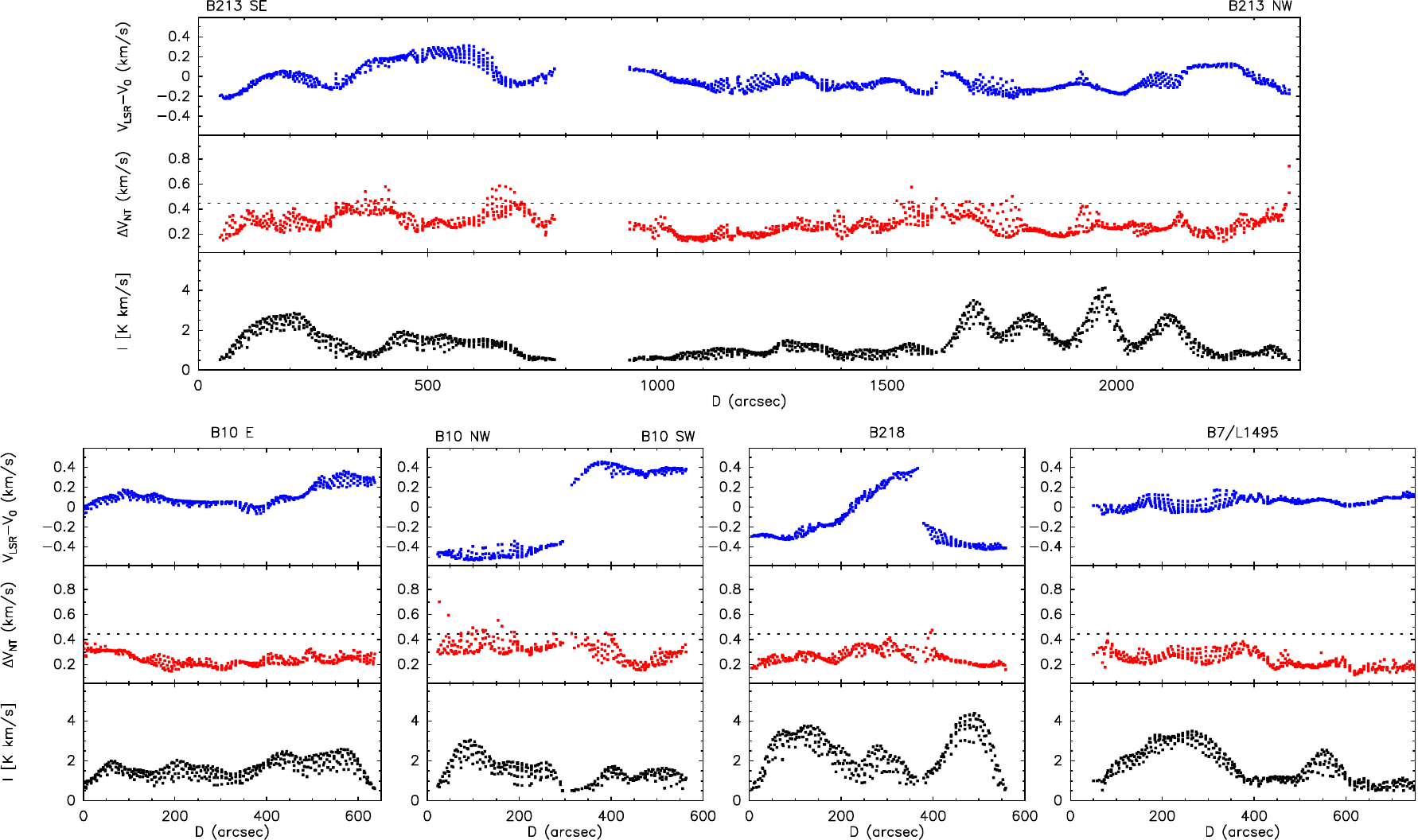}}
\caption{Velocity structure of the N$_2$H$^+$ emission along
the axis of the core chains. For each chain, the plot shows 
the velocity centroid in the top panel (blue symbols), the non-thermal
FWHM in the middle panel (red symbols), and the integrated intensity
in the bottom panel (black symbols). Only points within $30''$ of the
chain axis are shown to ensure proximity. 
The horizontal dashed line 
indicates the FWHM-equivalent of the sound velocity.
Note the smooth oscillations in the
velocity centroid and the predominance of subsonic
values in the non-thermal linewidth.
\label{oscillations}}
\end{figure*}

While the
maps in Fig.~\ref{n2hp_vel_maps}  
provide a good representation of the gas velocity field in two dimensions,
they provide limited
quantitative information on the gas kinematics.
This information is better appreciated in the
velocity profiles of 
Fig.~\ref{oscillations}. These profiles represent, as a function
of distance along the axis of each chain, the line center 
velocity (blue), the
non-thermal FWHM (red), and for reference, the
N$_2$H$^+$(1--0) integrated intensity (black).
To ensure physical proximity between the points,
the figure only shows data from positions within 30 arcsec 
from each chain axis. 
All panels use the same linear scale, and as a result, the
figure is dominated by the data from the 
B213 region, which has a length almost as large as the rest of
the chains combined.

\subsubsection{Line center velocity}
\label{sec_linecenter}

We first study the line-center
velocity, which is represented with blue squares in the top panels
of Fig.~\ref{oscillations}.
As can be seen, this parameter
presents little dispersion and
an almost oscillatory behavior in most panels.
In the longest B213 chain, 
the line center velocity oscillates repeatedly without
deviating 
by more than about 0.3~km~s$^{-1}$ from the mean
value over its full length of almost 1.5~pc.
The smaller
B10-E (eastern branch of B10) and B7/L1495 chains also show 
smooth velocity 
oscillations, again with close-to-constant mean values and
amplitudes of the order of 0.2~km~s$^{-1}$.

In contrast with the other chains, 
the western branch of B10 presents
a jump in velocity of about 1~km~s$^{-1}$ near
$D=300''$. This jump most likely results from
the presence in B10-W of two separate chains.
As discussed in Sect.~\ref{sect_c18o} and evidenced
by the double-peaked C$^{18}$O spectrum of Fig.~\ref{c18o_2vel}, 
two cloud components with velocities around 6 and 7~km~s$^{-1}$
coexist and overlap in B10-W.
The blue C$^{18}$O component lies mostly toward the NW, and
the red component lies mostly toward the
SW. This is also the distribution of the N$_2$H$^+$ 
line center velocities, which also match the velocities
of the two C$^{18}$O components. Since the N$_2$H$^+$ center velocity remains
almost constant toward each side of the jump, and the
jump coincides with a sharp drop of N$_2$H$^+$ emission 
(as shown in the 2D map of Fig.~\ref{n2hp_vel_maps}),
the most natural interpretation of the 
velocity jump in B10-W is that it
represents the transition
between the two C$^{18}$O components.
This implies that B10-W is not a single chain
of cores, but the chance alignment of
two different velocity components.
The recent numerical simulations      
of turbulent molecular clouds 
by \citet{moe14}
show that this type of chance alignment is
expected in regions like the L1495/B213 cloud.

Another discontinuity in the
velocity profiles of Fig.~\ref{oscillations}
occurs in B218 near $D=400''$.
As shown in \ref{n2hp_maps}, the B218 chain consists of 
three N$_2$H$^+$ cores, two of them starless and located
at each end of the chain
and a weaker one located near the center and associated with
the Class I object IRAS~04248+2612.
Fig.~\ref{oscillations} shows that the two starless cores 
have the same line center
velocity within 0.1~km~s$^{-1}$, but that
the region between them, where 
the N$_2$H$^+$ emission weakens and the IRAS source lies, 
presents a rapid shift in velocity towards the red.
This reddening of the N$_2$H$^+$ emission
coincides with the reddening of the C$^{18}$O emission
discussed in Sect.~\ref{sect_c18o}, and is highly localized toward the
vicinity of the YSO (see Fig.~\ref{n2hp_vel_maps}).
As discussed in Sect.~\ref{sect_c18o}
this reddening of the emission
likely results from the interaction
of the YSO with the surrounding cloud, and does not represent
a remnant of the pre-stellar motions in the chain.

The velocity oscillations in Fig.~\ref{oscillations} are similar to 
those found in L1517, also in the Taurus complex, by \citet{hac11}.
In L1517, several filaments 
presented a sinusoidal velocity pattern that
was approximately shifted by $\lambda/4$ from the also sinusoidal
pattern of column density. Such a shift
was interpreted as possible evidence of core-forming
gas motions along the filament axis, since analytic theory predicts
a $\lambda/4$ offset between density and velocity in the case of 
an unstable (core-forming) perturbation \citep{geh96,hac11}.
The core chains in L1495/B213 provide an ideal place to 
search for similar core-forming motions, since the multiplicity of cores
provides a strong constraint in the displacement between the velocity 
and column density profiles. The B213-NW chain, for example, 
contains 4 almost equally-spaced
cores, while the filaments in L1517 contained only two cores.
For this reason, we have fitted the N$_2$H$^+$
intensity profiles in Fig.~\ref{oscillations} with simple sinusoidal
functions (after subtracting a mean value), and we have compared the
velocity profiles with shifted versions of the intensity sinusoids.
While occasional matches for individual cores can be found, no chain presents
a systematic displacement between its velocity and intensity profiles
that applies to all cores and that
could be interpreted unambiguously as evidence of core-forming
motions along the chain axis. This 
lack of a systematic shift between the velocity and column density patterns
implies that the velocity oscillations in the chains
are not entirely due to the core-forming motions
predicted by the analytic theory. 

There are several possible origins
for the velocity oscillations seen in N$_2$H$^+$.
One possibility is that they still arise from core-forming motions, but that
the motions are more complex than what the simple analytic model assumes.
Most chains contain a mixture of starless cores and cores with embedded YSOs,
and this indicates that the contraction history of the chain 
must have been more irregular than what is assumed by
the simple model, in which all cores are formed simultaneously through 
the exponential growth of a single sinusoidal perturbation.
Another possibility is that the oscillations arise from motions that
pre exist the formation of the chains and the cores. \citet{hac13} found
that most C$^{18}$O filamentary components in the cloud present oscillations
in their velocity field irrespectively of their core-forming status.
In this case, the chains must start their evolution with an already perturbed
velocity field, and identifying any core-forming motions in such 
conditions may require a more complex analysis.
Numerical simulations of the formation and fragmentation of realistic
filamentary structures are needed to clarify this issue.

\subsubsection{Non-thermal velocity component}
\label{sect_nonth}

The non-thermal N$_2$H$^+$ component of the velocity field is represented with 
red squares in the middle panels of Fig.~\ref{oscillations}.
Like the line center velocity, the non-thermal component presents both
a smooth oscillatory behavior along each of the chains and 
a very low level of dispersion, except for a few
regions of moderate scatter.
The great majority of points lie below the sonic threshold
of 0.45~km~s$^{-1}$, indicating that subsonic
gas motions dominate the chains. 
For the data presented in the figure, which
consists of
points with separations of less than $30''$ from
the chain axis, the fraction of supersonic points is
only 2.1\%. Increasing the separation threshold to $90''$ 
adds more weak points, but only increases the
fraction of supersonic points to 3.7\%.
Subsonic points are therefore the norm, and supersonic
positions are rare.

In addition to being rare, the
supersonic points tend to lie in small groups,
implying that they result from localized causes.
Feedback from outflows is one of them,
as illustrated by the points in the vicinity of
IRAS 04169+2702, which lies in B213-NW, near $D = 1600''$.
This Class I source powers 
a well-known bipolar outflow previously studied in CO by
\citet{mor92}, \citet{bon96}, and \citet{nar12}.
As Fig.~\ref{n2hp_vel_maps} shows, the points of supersonic 
N$_2$H$^+$ linewidth lie north and south of IRAS 04169+2702
and approximately match
the bipolar distribution of the CO outflow lobes (see Fig.~10 in
\citealt{nar12}).

Other regions with supersonic N$_2$H$^+$ linewidths
of possible outflow origin are the vicinity of the three low-mass 
sources near $D = 400''$ in B213 SE,
which appears to have already been evacuated of dense gas,
and the already-mentioned vicinity of IRAS~04248+2612
in B218.
A few additional points with enhanced linewidth close to the supersonic
limit are associated with IRAS~04166+2706 in B213-NW near $D = 1950''$. 
This is another outflow
source of low luminosity that has evacuated a cavity in the surrounding 
dense core \citep{bon96,taf04b,san09}.

Not all regions with supersonic linewidth result from outflow feedback.
A group of supersonic points in B10-NW coincide
with the region of double-peaked C$^{18}$O spectra shown in Fig.~\ref{c18o_2vel},
and their enhanced linewidth likely results 
from this multi-component kinematics.
Indeed, some N$_2$H$^+$(1-0) spectra in this region show
two velocity components, and 
confusion between these components and the 
hyperfine structure seems to cause 
the larger linewidths seen in this region.
A similar mixing between components is likely to occur toward
the western part of B213-SE (D $\approx 650''$ in 
Fig.~\ref{oscillations}), where two velocity components at 6.7
and 7.0~km~s$^{-1}$ seem to coexist.

While we cannot exclude that some of the remaining
supersonic N$_2$H$^+$ linewidths correspond to positions with
intrinsically supersonic motions, the overwhelming dominance
of subsonic positions (>95\%) indicates that 
subsonic motions are a main characteristic of the chain gas.

\begin{figure}
\resizebox{\hsize}{!}{\includegraphics{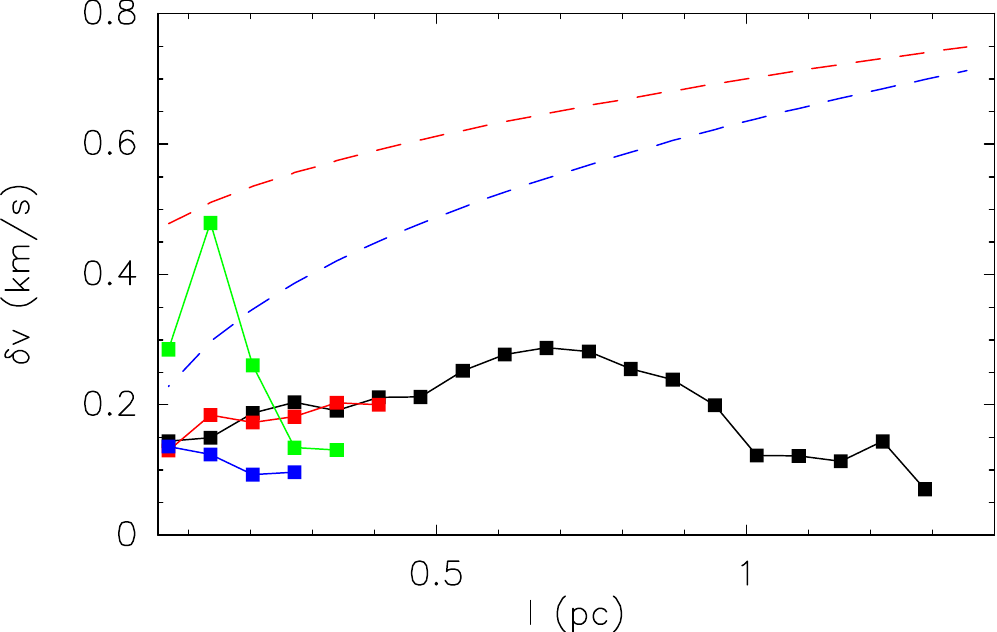}}
\caption{Structure function of the dense-gas velocity field.
The color-coded solid squares show the structure function as a function of
lag in B213 (black), B10 (red), B218 (green), and B7/L1495 (blue).
The blue dashed line represents the classical relation
from \citet{lar81}, and the red dashed line represents the
recently-determined core-velocity difference of \citet{qia12}.
\label{struct}}
\end{figure}

\subsubsection{Structure function}

\begin{figure*}
\resizebox{\hsize}{!}{\includegraphics{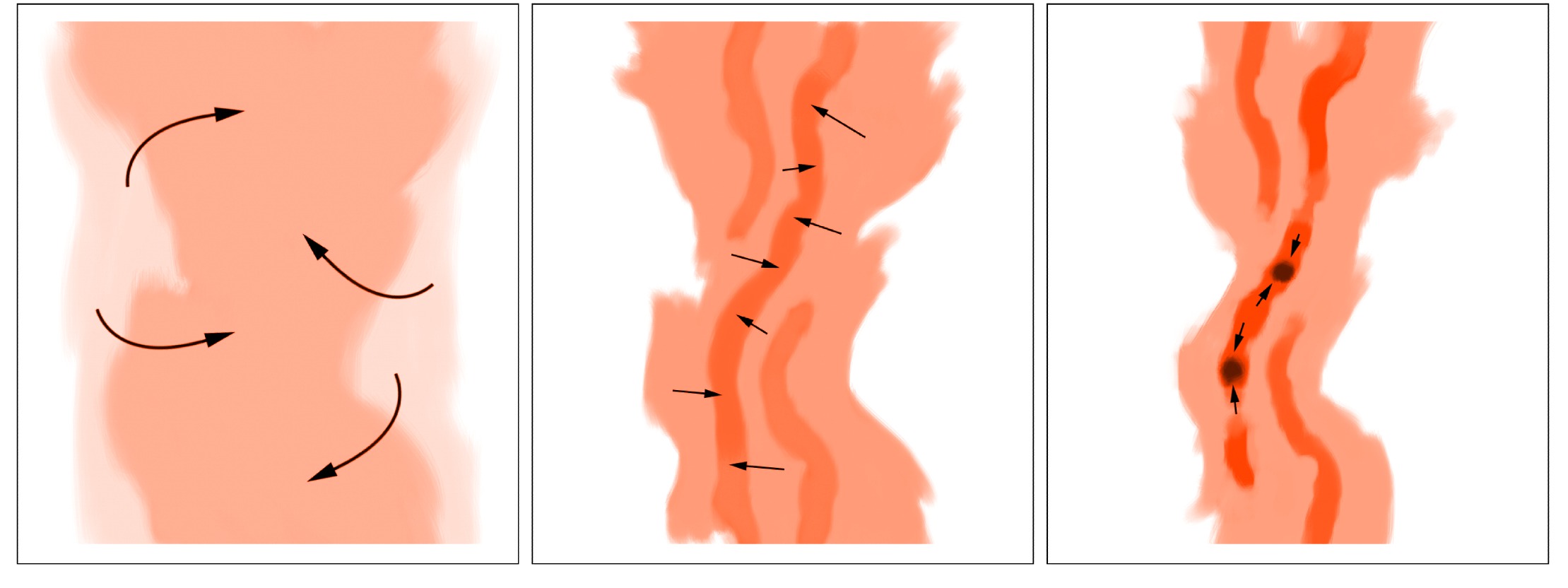}}
\caption{{\em Fray and fragment} scenario of core formation.
The three panels represent three different stages 
in the evolution of a large-scale filament like L1495/B213. 
In the first stage (left panel), two gas flows collide and
produce a filamentary density enhancement.
With time (middle panel), a
combination of residual turbulent motions and gravity
splits the gas into a series of intertwined velocity-coherent
fibers.  Finally (right panel), some fibers accumulate sufficient mass 
to reach the limit for gravitational fragmentation and form chains
of dense cores. 
\label{fray_fragment}}
\end{figure*}

The combination of subsonic motions and
large-scale continuity in the velocity field 
of the B213 gas filaments
led \citet{hac13} to refer to
these structures as ``velocity coherent'' (see also
\citealt{hac11}).
This term stresses the quiescent state of the filaments
compared to the 
cloud as a whole, which is characterized by 
a Kolmogoroff-type 
relation between size and velocity indicative of turbulent motions 
\citep{lar81}.
To further compare the kinematics of the gas in the chains
with that of the large-scale cloud,
we have carried out a structure-function analysis of the
N$_2$H$^+$ line-center velocities.
The structure function measures the mean difference 
in velocity between positions 
separated by a distance $l$, and is defined as
$$\delta v(l)^2 = \langle (v(x)-v(x+l))^2 \rangle,$$
where $v(x)$ is the gas velocity at an arbitrary 
position $x$ and the brackets
represent a spatial average over all positions separated by 
$l$. This function is commonly used as a descriptor
of the velocity field in a cloud, where 
it is systematically
found to depend on $l$ as a power law 
\citep{elm04,hey04}.

To estimate the structure function of the gas in the 
L1495/B213 chains, we have used the N$_2$H$^+$ line-center
velocities derived in Sect.~\ref{sec_n2hpkin} from hyperfine fits.
These velocities measure the radial component of the
velocity field, so our estimate of the structure function 
refers only to this radial component.
Fig.~\ref{struct} shows in colored squares
the structure function for 
all the chains but B10-W, which was found to be
the superposition of two different gas components.
As can be seen, the structure functions of 
B213, B10E, and B7/L1495 are approximately flat or present only
smooth oscillations as a function of $l$.
The structure function of 
B218 (green squares) presents a spike near
$l=0.15$~pc, but this feature is
caused by the strong reddening of the emission
near the YSO, which we saw is likely due to outflow feedback.
At larger distances, corresponding to the separation between the starless 
cores at each side of the YSO, the structure function of B218 converges
to a low value comparable to that of the other chains.

The flat structure functions of the chains 
contrast with the power-law functions
of the extended gas indicative of turbulent motions
\citep{lar81,elm04,hey04}.
This is illustrated in Fig.~\ref{struct} with two dashed lines.
The blue line represents the classical \citet{lar81} relation,
and the red line is the
structure function estimated by \citet{qia12}
for Taurus using a large-scale map of the $^{13}$CO emission
\citep{gol08}. As seen in the figure, both
dashed lines follow a similar power-law increase with
$l$ and 
deviate systematically from the flat structure functions of the
chains.
While the deviation between chains and cloud
is most prominent in the
B213 chain due to its larger length, it is 
also significant in the other three chains 
due to the consistency of their behavior.

The flatness of the structure functions in Fig.~\ref{struct} adds
evidence to the suggestion by \citet{hac13}, from C$^{18}$O data,
that the gas in the velocity-coherent filaments 
has decoupled from the turbulent velocity field that dominates the
cloud as a whole. The observations, therefore, imply that turbulence
does not dissipate at the $\approx 0.1$~pc 
scale of the dense cores \citep{goo98}, but at a larger
scale of approximately 0.5~pc that corresponds to the C$^{18}$O
velocity-coherent fibers.
As a result of this dissipation of the turbulence at larger scales,
the condensation of the cores out of the chain gas
must involve little kinematic change, which is in agreement with
the smooth oscillations seen in the radial profiles of velocity.
Thus, while turbulence dominates the motions of the cloud gas at 
large scales, it appears to have been dissipated before the gas
condenses into close-to-spherical dense cores.

\section{Implications for core and star formation}

We now combine our analysis of the internal structure of the
N$_2$H$^+$ chains with the study
of the large-scale C$^{18}$O emission by \citet{hac13}
and a number of arguments from analytic theory and 
numerical simulations to present a
scenario of core and star formation in
the L1495/B213 region. 
A schematic view of this scenario is shown in Fig.~\ref{fray_fragment} with
a three-step time sequence of the gas evolution in the
cloud. For reasons that should be clear below, we refer to this 
scenario as {\em ``fray and fragment.''}

The first step in the scenario consists of
the formation of 
the 10~pc-long L1495/B213 region. This event, like the formation
of the rest of the Taurus molecular cloud, most likely resulted from
the collision between two large-scale flows of gas, as proposed
by a number authors and implied by numerical 
simulations \citep{bal99,har01,pad02,mac04,vaz07,hei08}.
A lower limit to the velocity of this collision can be estimated using
the velocity spread of the gas in the L1495/B213 filament.
The C$^{18}$O data of \citet{hac13} show some 
spectra containing multiple velocity components that
range in LSR centroid velocity between 4.8 and 7.0 km~s$^{-1}$
(their Fig.~8). This spread implies that 
the relative velocity of the converging flows 
was at least 2.2~km~s$^{-1}$.

Since the velocity spread of 2.2~km~s$^{-1}$ refers to C$^{18}$O-emitting gas, 
which has a typical temperature of 10~K, it corresponds 
to a Mach number of about 11. This implies that the 
internal motions in the large-scale
filament of L1495/B213 definitely belong to the supersonic regime.

The second step in the scenario of Fig.~\ref{fray_fragment} is the generation
of substructure inside the 10~pc-long L1495/B213 filamentary cloud.
This substructure was studied in detail by \citet{hac13}, who
analyzed the C$^{18}$O emission from the cloud
with the Friends In VElocity (FIVE)
algorithm and identified 35 distinct velocity-coherent
fibers.
These fibers run approximately parallel
to the 10~pc-long L1495/B213 cloud and 
criss-cross each other like the threads of a frayed rope.

While the relative velocity between fibers spans the full
2.2~km~s$^{-1}$ range, the velocity dispersion
of the gas inside each fibers
is much lower and lies in the subsonic or slightly
transonic regime. 
Such combination of low internal velocity dispersion and 
large-scale continuity of the velocity implies 
that these fibers have already decoupled
from the large-scale turbulent velocity field of the cloud,
and that therefore represent a critical scale of velocity dissipation.
The recent numerical simulations of
cloud turbulence by \citet{moe14}
seem to confirm this. The simulations
show that large-scale filamentary structures quickly 
evolve into bundles of intertwined fibers
like those of L1495/B213
due to the combined effect of vorticity and self gravity
(see also \citealt{kri13}, \citealt{smi13}, and \citealt{mye14}).

The final step of the scenario shown in Fig.~\ref{fray_fragment}
is the formation
of dense cores out of the velocity-coherent fibers.
\citet{hac13} showed that only a few of these
fibers are responsible for the formation
of all dense cores cores in L1495/B213,
and referred to them as ``fertile.'' 
These fertile fibers preferentially form multiple cores, 
so they seem to have an intrinsic predisposition
to core formation along most of their length.
A natural origin for this predisposition is 
gravitational instability, as implied by the
higher value of the 
mass-per-unit-length in these fibers.

\begin{figure}
\resizebox{\hsize}{!}{\includegraphics{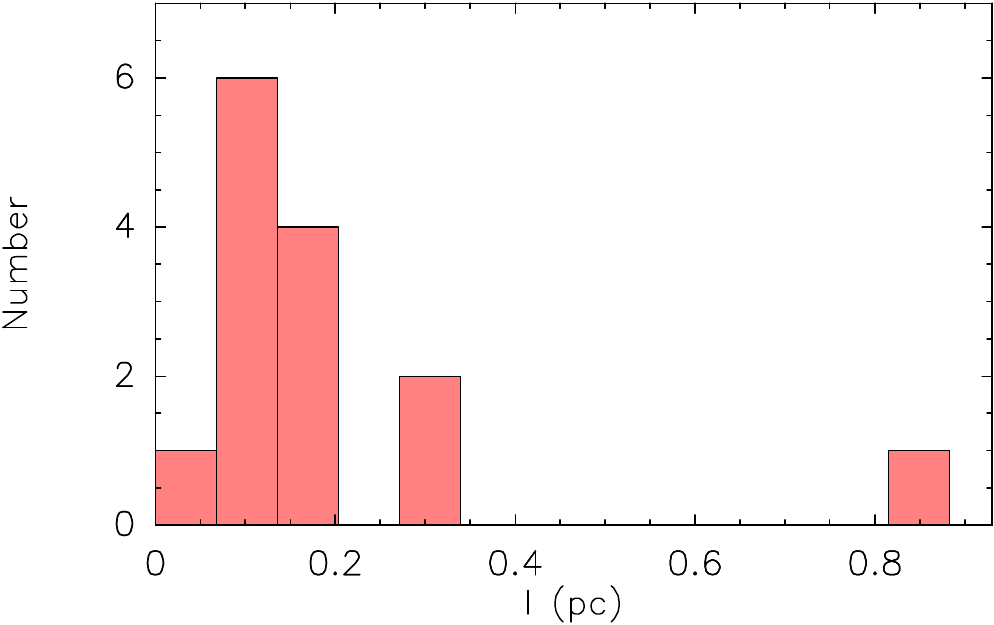}}
\caption{Distance to the nearest neighbor among
the N$_2$H$^+$ cores in L1495/B213 (no projection correction applied).
Note how most values lie below 0.2~pc.
\label{nearest}}
\end{figure}

To test this interpretation,  we have studied the typical separation 
between dense cores along the chains. This has been done
using the core positions determined in Sect.~2 and calculating
the distribution of distance
to the nearest neighbor.
Fig.~\ref{nearest}
shows a histogram with the results. As can be seen,
the distribution is dominated by values smaller
than 0.2~pc, with 3/4 of the points lying
below this length. The median of the distribution
is 0.14~pc and the mean is 0.2~pc, although this latter value
is biased due to the the large contribution from
the isolated core in B211.

Since the observed distances represent measures in 
the plane of the sky,
they have to be corrected for inclination.
As discussed in Sect.~\ref{sect_radprof}, the inclination angle 
of the chains is unknown, but it is unlikely to be
very large due to the already long appearance of
the L1495/B213 complex. Thus, we have
assumed a moderate inclination angle
of 45 degrees, which implies 
a foreshortening factor of only 1.4. With this correction,
the true median distance between cores is estimated to be
approximately 0.2~pc.

To compare the above inter-core distance with the expectation from
gravitational fragmentation, we use as a guide the 
self-gravitating isothermal cylinder model.
This model can only represent a first-order approximation to the chains,
since we have seen that their mass per unit length
likely exceed the model prediction.
In addition, the turbulent origin of the fibers and chains
makes it unlikely that the gas in them has had sufficient time to fully 
relax
to an isothermal equilibrium configuration, or even to acquire the 
axial symmetry required for a cylindrical geometry.
Still, the deviations from equilibrium are unlikely to be very large,
since the prevalence of subsonic linewidths
in the spectra indicates an absence of shocks.
Under these conditions,
the isothermal cylinder model must provide a reasonable
approximation, especially considering that
the characteristics of gravitational fragmentation are 
robust with respect to changes in geometry and
contributions from rotation and magnetic fields
\citep{lar85}.

As shown by \citet{sto63},
an isothermal cylinder 
with an equilibrium mass per unit length 
is unstable to
sinusoidal perturbations that have a
wavelength larger than a critical value of
$3.14\; c/\sqrt{G \rho_0},$
where $\rho_c$ is the central density of the filament.
These perturbations fragment the cylinder into a
a series of equally-spaced condensations
whose separation 
is given by the wavelength of the perturbation
applied to the system, or
if no single-wavelength perturbation is imposed, 
by the wavelength of
the fastest-growing unstable mode, which is
twice the critical wavelength
(\citealt{nag87}, also \citealt{lar85,inu97,fis12}).
In Sect.~\ref{sect_radprof} we estimated that the core chains have a typical 
pre-fragmentation central density of $\approx 0.65 \times 10^5$~cm$^{-3}$.
For the isothermal cylinder model, this density
implies a critical wavelength of 0.14~pc, and a
fastest-growing mode wavelength of 0.28~pc,
assuming the gas is at 10~K. 
These values are in reasonable agreement with our
estimate of a typical inter-core distance of 0.2~pc.
Strictly speaking, 
an inter-core value closer to 0.3~pc
would have been expected, but it is also possible that
the initial central density of the fibers 
has been underestimated, and that this has led to an
overestimate of the
expected critical wavelength and
the separation between the cores. Given 
all the uncertainties in our estimates of the
chain physical parameters, it appears that
gravitational fragmentation is
the likely mechanism responsible for the 
formation of the
dense cores inside the chains.

If correct, the picture of core formation by 
gravitational fragmentation 
has several important consequences.
In first place, fragmentation can only occur if
an isothermal cylinder has sufficient
mass per unit length. 
Since core formation in L1495/B213 is restricted to a
small number of fertile fibers, it appears that
only fertile fibers reach the required
mass-per-unit-length, and that the rest of the fibers 
remain sterile because they never accumulate enough mass
(this is supported by the 
analysis of \citealt{hac13}).
This interpretation suggests that core formation in L1495/B213
is regulated by how much mass the fibers can 
accumulate, and that inability to reach the critical
mass is a main bottleneck in the core and star formation process.
Given the turbulent state of the large-scale cloud,
early dissipation of sterile fibers by shocks is the
most likely mechanism to limit their growth.
Star formation is thus limited, not by failed cores, but by
failed (or sterile) fibers.

Another consequence of the gravitational fragmentation scenario
is that dense cores may not be equilibrium structures, since
they originate from an instability that cannot be reversed
without external energy injection.
This conclusion may seem to 
contradict the idea that cores are in gravitational equilibrium,
as implied by the good fit of their radial density profiles
with models of isothermal (Bonnor-Ebert) equilibrium \citep{alv01}.
Equilibrium density profiles, however, are not a guarantee
of true hydrostatic equilibrium, since they are expected 
to occur also
during the first stages of gravitational collapse \citep{kan05}.
Evidence of (subsonic) gas contraction in starless cores
is in fact often found when combining observations
of optically thick and thin tracers \citep{lee99}.

While cylindrical models represent useful tools
to understand core formation
in the chains, they can
only represent a first order approximation. 
As seen in the N$_2$H$^+$ maps, the chains have a typical full length of 0.5~pc,
which is only about twice the wavelength of the fastest-growing
mode. The chains therefore hardly qualify as ``infinitely long,'' and 
their fragmentation is likely to have been affected by edge effects.
Chain edges, however,
are unlikely to be sharp, since this would favor the rapid production of
condensations near the boundaries \citep{bur04}, which is
not seen in the maps.
More likely, the edges of the chains 
are characterized by a smooth density decrease,
which is expected to favor fragmentation in the chain interior
\citep{nel93}.
Other deviations from the idealized infinite cylinder model include
the natural bending of some of the chains, like B10, and the
likely presence of significant initial perturbations in the density 
profile as a result of their turbulent environment.
These additional elements are
likely responsible for the already-discussed irregular 
pattern of fragmentation in the chains, by which cores that have 
already formed stars are intermixed with starless cores
at an earlier stage of evolution. Numerical simulations
of bundles of filaments, like those presented by 
\citet{moe14}, are needed to explore how 
gravitational fragmentation occurs under 
more realistic conditions.

We finish by noting that the
{\em fray and fragment} scenario proposed here has been
inspired by the analysis of L1495/B213, but that it could 
apply to other regions of core and star formation.
The
analysis of the N$_2$H$^+$ emission from the NGC~1333 cluster-forming
region by Hacar et al. (2014, in preparation), for example, shows
that the dense gas in this region forms a network of velocity-coherent
fibers very similar to those found in L1495/B213, although with 
a higher density of fibers per unit volume, as it would be
expected for a cluster-forming region.
In addition,
a number of large-scale dust-continuum images of clouds
show that what initially appears as a
single large-scale filament, on close inspection is resolved into a network
of closely-aligned, small-scale fibers containing 
dense cores and embedded young stars. 
This is noticeable in some of the SPIRE images of the
Herschel Gould Belt Survey, and is
especially prominent in the APEX image
of Orion presented by the ESO press release 
{\em eso1321}\footnote{\url{http://www.eso.org/public/news/eso1321/}}
(see also \citealt{sta13}). 
Velocity-resolved observations of these filamentary structures
are needed to test whether a {\em fray and fragment} scenario
can explain core formation in other regions and in
more massive environments.

\section{Conclusions}

We have presented new N$_2$H$^+$(1--0) and C$^{18}$O(2--1) observations 
of the dense regions of L1498/B213 in Taurus made with the IRAM 30m telescope
and complemented with archival dust-continuum observations
from the Herschel Space Observatory. From the analysis of 
the combined dataset, we have reached the following main conclusions.

1. The dense cores in L1498/B213 tend to cluster in linear
groups of three cores on average, which we call chains. This clustering is
evident in the maps of integrated emission, and produces
a significant peak in the mean surface density of companions
for separations smaller than about $500''$ ($\approx 0.3$~pc). 
Only 3 cores out of 22 have no neighbor within $300''$ (0.2~pc).

2. The N$_2$H$^+$ integrated intensities are correlated with
the H$_2$ column densities estimated from the SPIRE-Herschel data.
This correlation is approximately linear, and in some chains
it has a small threshold of H$_2$ column
density that seems to arise from
additional gas components. 
The linear correlation indicates that towards the chains
the N$_2$H$^+$ emission 
traces most of the material seen in the Herschel images, and therefore
provides velocity information to the continuum data.
It also indicates that the N$_2$H$^+$ abundance 
is approximately uniform with a value of
$5 \times 10^{-10}$.

3. A simplified Monte Carlo model of the N$_2$H$^+$
radial profiles indicates that the density of the
chains decreases 
with radius as a softened power law. Typical central densities 
are $6-7 \times 10^4$~cm$^{-3}$ and half-maximum
diameters are 0.05~pc.

4. While the C$^{18}$O emission reveals the presence 
of multiple velocity components separated by supersonic speeds 
(analyzed in detail by \citealt{hac13}).
the N$_2$H$^+$ emission shows that the 
dense gas is overwhelmingly subsonic and continuous
in velocity. This gas has appears to have decoupled from
the turbulent velocity field of the cloud and 
does not follow the standard Larson relation 
despite extending for up to 1~pc in length.

5. When combined with the analysis of the C$^{18}$O emission
from \citet{hac13}, our observations 
suggest a scenario of core formation which we refer to 
as {\em fray and fragment}. In this scenario,
a collision between two supersonic gas flows
has created the large-scale L1498/B213 filament.
Due to a combination of turbulence and  self-gravity,
the large-scale filament has 
split into a network of smaller and intertwined
filamentary structures or fibers ({\em fray} step).
Some of these fibers have accumulated
enough material to exceed the mass-per-unit-length
limit of gravitational instability and to
{\em fragment} forming chains of dense cores.
Although this scenario is motivated by
L1498/B213, additional observations suggest
that it may apply to other regions.

\begin{acknowledgements}
 We thank the IRAM staff for support during the observations, and
 Markus Schmalzl for generously providing us with 
 his extinction data and for information
 on the relation between NIR extinction and SPIRE $500 \mu$m
 emission.
 We also thank Joaqu\'{\i}n Santiago-Garc\'{\i}a for
 his rendering of the {\em fray and fragment} scenario 
 shown in Fig.~\ref{fray_fragment}.
 This research was performed in part thanks to financial support from projects 
 FIS2012-32096 and AYA2012-32032 of Spanish MINECO and from the MICINN program
 CONSOLIDER INGENIO 2010, grant ``Molecular Astrophysics:
 The Herschel and ALMA era - ASTROMOL'' (ref.: CSD2009-00038).
 AH acknowledges support from the Austrian Science Fund (FWF).
 This research has made use of NASA's Astrophysics Data System
 Bibliographic Services together with the SIMBAD database and
 the VizieR catalogue access tool
 operated at CDS, Strasbourg, France.
\end{acknowledgements}



\begin{thebibliography}{}
\bibitem[Aikawa et al.(2005)]{aik05} Aikawa, Y., Herbst, E., 
Roberts, H., \& Caselli, P.\ 2005, \apj, 620, 330 
\bibitem[Alves et al.(2001)]{alv01} Alves, J.~F., Lada, 
C.~J., \& Lada, E.~A.\ 2001, \nat, 409, 159 
\bibitem[Andr{\'e} et al.(2010)]{and10} Andr{\'e}, P., Men'shchikov, A.,
Bontemps, S., et al.\ 2010, \aap, 518, L102
\bibitem[Andr{\'e} et al.(2013)]{and13} Andr{\'e}, P., Di 
Francesco, J., Ward-Thompson, D., et al.\ 2013, arXiv:1312.6232 
\bibitem[Arzoumanian et al.(2011)]{arz11} Arzoumanian, D., Andr{\'e}, P., 
Didelon, P., et al.\ 2011, \aap, 529, L6
\bibitem[Ballesteros-Paredes et al.(1999)]{bal99} Ballesteros-Paredes, 
J., V{\'a}zquez-Semadeni, E., \& Scalo, J.\ 1999, \apj, 515, 286 
\bibitem[Barnard(1907)]{bar07} Barnard, E.~E.\ 1907, \apj, 25, 218 
\bibitem[Benson \& Myers(1989)]{ben89} Benson, P.~J., \& Myers, P.~C.\ 1989, 
\apjs, 71, 89
\bibitem[Bergin et al.(2002)]{ber02} Bergin, E.~A., Alves,
J., Huard, T., \& Lada, C.~J.\ 2002, \apjl, 570, L101
\bibitem[Bergin \& Tafalla(2007)]{ber07} Bergin, E.~A., \& Tafalla, M.\ 2007, \araa, 45, 339
\bibitem[Bernes(1979)]{ber79} Bernes, C.\ 1979, \aap, 73, 67 
\bibitem[Bontemps et al.(1996)]{bon96} Bontemps, S., Andre, P., Terebey, S., 
\& Cabrit, S.\ 1996, \aap, 311, 858 
\bibitem[Burkert \& Hartmann(2004)]{bur04} Burkert, A., \& Hartmann, L.\ 2004, 
\apj, 616, 288 
\bibitem[Carter et al.(2012)]{car12} Carter, M., Lazareff, B., Maier, D., 
et al.\ 2012, \aap, 538, A89 
\bibitem[Caselli et al.(2002)]{cas02} Caselli, P., Benson,
 P.~J., Myers, P.~C., \& Tafalla, M.\ 2002, \apj, 572, 238
 \bibitem[Caselli et al.(1995)]{cas95} Caselli, P., Myers, 
 P.~C., \& Thaddeus, P.\ 1995, \apjl, 455, L77 
 \bibitem[Caselli et al.(1999)]{cas99} Caselli, P., Walmsley,
 C.~M., Tafalla, M., Dore, L., \& Myers, P.~C.\ 1999, \apjl, 523, L165
\bibitem[Cernicharo et al.(1985)]{cer85} Cernicharo, J., Bachiller, R.,
\& Duvert, G.\ 1985, \aap, 149, 273
\bibitem[Daniel et al.(2005)]{dan05} Daniel, F., Dubernet, 
M.-L., Meuwly, M., Cernicharo, J., \& Pagani, L.\ 2005, \mnras, 363, 1083 
\bibitem[di Francesco et al.(2007)]{dif07} di Francesco, J., 
Evans, N.~J., II, Caselli, P., et al.\ 2007, Protostars and Planets V, 17
\bibitem[Dobashi et al.(2005)]{dob05} Dobashi, K., Uehara, 
H., Kandori, R., et al.\ 2005, \pasj, 57, 1 
\bibitem[Elias(1978)]{eli78} Elias, J.~H.\ 1978, \apj, 224, 857
\bibitem[Elmegreen \& Scalo(2004)]{elm04} Elmegreen, B.~G., \& Scalo, 
J.\ 2004, \araa, 42, 211 
\bibitem[Evans et al.(2009)]{eva09} Evans, N.~J., II, Dunham,
M.~M., J{\o}rgensen, J.~K., et al.\ 2009, \apjs, 181, 321
\bibitem[Evans et al.(2001)]{eva01} Evans, N.~J., II, 
Rawlings, J.~M.~C., Shirley, Y.~L., \& Mundy, L.~G.\ 2001, \apj, 557, 193 
\bibitem[Fischera \& Martin(2012)]{fis12} Fischera, J., \& Martin, P.~G.\ 
2012, \aap, 542, A77
\bibitem[Gaida et al.(1984)]{gai84} Gaida, M., Ungerechts, H., 
\& Winnewisser, G.\ 1984, \aap, 137, 17
\bibitem[Galli et al.(2002)]{gal02} Galli, D., Walmsley, M., 
\& Gon{\c c}alves, J.\ 2002, \aap, 394, 275 
\bibitem[Gehman et al.(1996)]{geh96} Gehman, C.~S., Adams, 
F.~C., \& Watkins, R.\ 1996, \apj, 472, 673 
\bibitem[Goldsmith et al.(2008)]{gol08} Goldsmith, P.~F.,
Heyer, M., Narayanan, G., et al.\ 2008, \apj, 680, 428
\bibitem[Gomez et al.(1993)]{gom93} Gomez, M., Hartmann, L., 
Kenyon, S.~J., \& Hewett, R.\ 1993, \aj, 105, 1927 
\bibitem[Gomez et al.(1997)]{gom97} Gomez, M., Whitney, 
B.~A., \& Kenyon, S.~J.\ 1997, \aj, 114, 1138 
\bibitem[Goodman et al.(1998)]{goo98} Goodman, A.~A., 
Barranco, J.~A., Wilner, D.~J., \& Heyer, M.~H.\ 1998, \apj, 504, 223 
\bibitem[Griffin et al.(2010)]{grif10} Griffin, M.~J., Abergel, A., Abreu, A., 
et al.\ 2010, \aap, 518, L3 
\bibitem[Hacar \& Tafalla(2011)]{hac11} Hacar, A., \& Tafalla, M.\ 2011,
\aap, 533, A34
\bibitem[Hacar et al.(2013)]{hac13} Hacar, A., Tafalla, M., Kauffmann, J., 
\& Kov{\'a}cs, A.\ 2013, \aap, 554, A55 
\bibitem[Hartmann et al.(2001)]{har01} Hartmann, L., 
Ballesteros-Paredes, J., \& Bergin, E.~A.\ 2001, \apj, 562, 852 
\bibitem[Hartmann(2002)]{har02} Hartmann, L.\ 2002, \apj, 578, 914
\bibitem[Heitsch et al.(2008)]{hei08} Heitsch, F., Hartmann, L.~W., Slyz, 
A.~D., Devriendt, J.~E.~G., \& Burkert, A.\ 2008, \apj, 674, 316 
\bibitem[Hennemann et al.(2012)]{hen12} Hennemann, M., Motte, F., 
Schneider, N., et al.\ 2012, \aap, 543, L3 
\bibitem[Henning et al.(2010)]{hen10} Henning, T., Linz, H., Krause, O., 
et al.\ 2010, \aap, 518, L95 
\bibitem[Heyer \& Brunt(2004)]{hey04} Heyer, M.~H., \& Brunt, C.~M.\ 2004, 
\apjl, 615, L45 
\bibitem[Hildebrand(1983)]{hil83} Hildebrand, R.~H.\ 1983, \qjras, 24, 267 
\bibitem[Inutsuka \& Miyama(1997)]{inu97} Inutsuka, S.-i., \& Miyama, S.~M.\ 
1997, \apj, 480, 681 
\bibitem[Jijina et al.(1999)]{jij99} Jijina, J., Myers, 
P.~C., \& Adams, F.~C.\ 1999, \apjs, 125, 161 
\bibitem[Juvela et al.(2011)]{juv11} Juvela, M., Ristorcelli, I., 
Pelkonen, V.-M., et al.\ 2011, \aap, 527, A111 
\bibitem[Kandori et al.(2005)]{kan05} Kandori, R., Nakajima, 
Y., Tamura, M., et al.\ 2005, \aj, 130, 2166 
\bibitem[Kirk et al.(2013)]{kir13} Kirk, J.~M., 
Ward-Thompson, D., Palmeirim, P., et al.\ 2013, \mnras, 432, 1424 
\bibitem[K{\"o}nyves et al.(2010)]{kon10} K{\"o}nyves, V., Andr{\'e}, P., 
Men'shchikov, A., et al.\ 2010, \aap, 518, L106
\bibitem[Kritsuk et al.(2013)]{kri13} Kritsuk, A.~G., Lee, 
C.~T., \& Norman, M.~L.\ 2013, \mnras, 436, 3247 
\bibitem[Larson(1981)]{lar81} Larson, R.~B.\ 1981, \mnras, 194, 809
\bibitem[Larson(1985)]{lar85} Larson, R.~B.\ 1985, \mnras, 214, 379 
\bibitem[Larson(1995)]{lar95} Larson, R.~B.\ 1995, \mnras, 272, 213
\bibitem[Launhardt et al.(2013)]{lau13} Launhardt, R., Stutz, A.~M., 
Schmiedeke, A., et al.\ 2013, \aap, 551, A98 
\bibitem[Lee et al.(1999)]{lee99} Lee, C.~W., Myers, P.~C., 
\& Tafalla, M.\ 1999, \apj, 526, 788 
\bibitem[Lee et al.(2001)]{lee01} Lee, C.~W., Myers, P.~C., 
\& Tafalla, M.\ 2001, \apjs, 136, 703 
\bibitem[Lombardi et al.(2010)]{lom10} Lombardi, M., Lada, C.~J., \& 
Alves, J.\ 2010, \aap, 512, A67 
\bibitem[Luhman et al.(2010)]{luh10} Luhman, K.~L., Allen, 
P.~R., Espaillat, C., Hartmann, L., \& Calvet, N.\ 2010, \apjs, 186, 111 
\bibitem[Mac Low \& Klessen(2004)]{mac04} Mac Low, M.-M., \& Klessen, R.~S.\ 
2004, Reviews of Modern Physics, 76, 125 
\bibitem[Mangum et al.(2007)]{man07} Mangum, J.~G., Emerson, D.~T., 
\& Greisen, E.~W.\ 2007, \aap, 474, 679 
\bibitem[Moeckel \& Burkert(2014)]{moe14} Moeckel, N., \& Burkert, A.\ 2014, 
arXiv:1402.261
\bibitem[Molinari et al.(2010)]{mol10} Molinari, S., Swinyard, B., Bally, J., et al.\ 2010,
\bibitem[Mooley et al.(2013)]{moo13} Mooley, K., Hillenbrand, 
L., Rebull, L., Padgett, D., \& Knapp, G.\ 2013, \apj, 771, 110
\aap, 518, L100
\bibitem[Moriarty-Schieven et al.(1992)]{mor92}
Moriarty-Schieven, G.~H., Wannier, P.~G., Tamura, M.,
\& Keene, J.\ 1992, \apj, 400, 260
\bibitem[Myers(2009)]{mye09} Myers, P.~C.\ 2009, \apj, 700, 1609
\bibitem[Myers et al.(2014)]{mye14} Myers, A.~T., Klein, 
R.~I., Krumholz, M.~R., \& McKee, C.~F.\ 2014, \mnras, 439, 3420 
\bibitem[Nagasawa(1987)]{nag87} Nagasawa, M.\ 1987, Progress 
of Theoretical Physics, 77, 635 
\bibitem[Nakamura et al.(1993)]{nak93} Nakamura, F., Hanawa, 
T., \& Nakano, T.\ 1993, \pasj, 45, 551 
\bibitem[Narayanan et al.(2012)]{nar12} Narayanan, G., Snell, 
R., \& Bemis, A.\ 2012, \mnras, 425, 2641 
\bibitem[Nelson \& Papaloizou(1993)]{nel93} Nelson, R.~P., \& Papaloizou, 
J.~C.~B.\ 1993, \mnras, 265, 905
\bibitem[Ostriker(1964)]{ost64} Ostriker, J.\ 1964, \apj,
140, 1056
\bibitem[Padoan \& Nordlund(2002)]{pad02} Padoan, P., \& Nordlund, {\AA}.\ 
2002, \apj, 576, 870 
\bibitem[Palmeirim et al.(2013)]{pal13} Palmeirim, P., Andr{\'e}, P., 
Kirk, J., et al.\ 2013, \aap, 550, A38 
\bibitem[Pilbratt et al.(2010)]{pil10} Pilbratt, G.~L., et al.\ 2010, \aap, 518, L1 
\bibitem[Qian et al.(2012)]{qia12} Qian, L., Li, D., 
\& Goldsmith, P.~F.\ 2012, \apj, 760, 147 
\bibitem[Rebull et al.(2010)]{reb10} Rebull, L.~M., Padgett, 
D.~L., McCabe, C.-E., et al.\ 2010, \apjs, 186, 259 
\bibitem[Recchi et al.(2013)]{rec13} Recchi, S., Hacar, A., \& Palestini, 
A.\ 2013, \aap, 558, A27 
\bibitem[Sandell et al.(2011)]{san11} Sandell, G., Weintraub, 
D.~A., \& Hamidouche, M.\ 2011, \apj, 727, 26
\bibitem[Santiago-Garc{\'{\i}}a et al.(2009)]{san09} Santiago-Garc{\'{\i}}a, J., 
Tafalla, M., Johnstone, D., \& Bachiller, R.\ 2009, \aap, 495, 169 
\bibitem[Schmalzl et al.(2010)]{sch10} Schmalzl, M.,
Kainulainen, J., Quanz, S.~P., et al.\ 2010, \apj, 725, 1327
\bibitem[Schneider \& Elmegreen(1979)]{sch79} Schneider, S., \& Elmegreen, 
B.~G.\ 1979, \apjs, 41, 87
\bibitem[Sch{\"o}ier et al.(2005)]{sch05} Sch{\"o}ier, F.~L., van der Tak, 
F.~F.~S., van Dishoeck, E.~F., \& Black, J.~H.\ 2005, \aap, 432, 369 
\bibitem[Simon(1997)]{sim97} Simon, M.\ 1997, \apjl, 482, L81
\bibitem[Smith et al.(2013)]{smi13} Smith, R.~J., Shetty, R., 
Beuther, H., Klessen, R.~S., \& Bonnell, I.~A.\ 2013, \apj, 771, 24 
\bibitem[Stanke et al.(2013)]{sta13} Stanke, T., Stutz, A., 
Megeath, T., \& HOPS Team 2013, Protostars and Planets VI Posters, 4 
\bibitem[Stod{\'o}lkiewicz(1963)]{sto63} Stod{\'o}lkiewicz,
J.~S.\ 1963, Acta Astronomica, 13, 30
\bibitem[Suutarinen et al.(2013)]{suu13} Suutarinen, A., Haikala, L.~K., 
Harju, J., et al.\ 2013, \aap, 555, A140 
\bibitem[Tafalla et al.(2002)]{taf02} Tafalla, M., Myers,
P.~C., Caselli, P., Walmsley, C.~M., \& Comito, C.\ 2002, \apj, 569, 815
\bibitem[Tafalla et al.(2004a)]{taf04a} Tafalla, M., Myers, P.~C., Caselli, P., 
\& Walmsley, C.~M.\ 2004, \aap, 416, 191
\bibitem[Tafalla et al.(2004b)]{taf04b} Tafalla, M., Santiago, J., Johnstone, D., 
\& Bachiller, R.\ 2004, \aap, 423, L21 
\bibitem[Tatematsu et al.(2004)]{tat04} Tatematsu, K., 
Umemoto, T., Kandori, R., \& Sekimoto, Y.\ 2004, \apj, 606, 333 
\bibitem[V{\'a}zquez-Semadeni et al.(2007)]{vaz07} V{\'a}zquez-Semadeni, E., 
G{\'o}mez, G.~C., Jappsen, A.~K., et al.\ 2007, \apj, 657, 870 
\bibitem[Ward-Thompson et al.(2007)]{war07} Ward-Thompson,
D., Andr{\'e}, P., Crutcher, R., et al.\ 2007, Protostars and Planets V, 33
\bibitem[Ysard et al.(2013)]{ysa13} Ysard, N., Abergel, A., Ristorcelli, I., 
et al.\ 2013, \aap, 559, A133 



\end{thebibliography}
\end{document}